\documentclass[reprint,floatfix, twocolumn]{revtex4-1}

\usepackage{pifont} 

\usepackage[utf8]{inputenc}
\usepackage[a4paper]{geometry}
\usepackage{hyperref}
\usepackage{graphicx}
\usepackage{xcolor}
\usepackage{multirow}
\usepackage{ulem}

\usepackage{setspace}
\usepackage[caption=false]{subfig}
\usepackage[T1]{fontenc}

\usepackage[charter]{mathdesign}
\usepackage{amsmath}
\usepackage{amsfonts, relsize, color}
\usepackage{mathrsfs}
\usepackage{mathtools}
\usepackage{physics}
\usepackage{bbm}

\newcommand{\ncmd}{\newcommand}
\ncmd{\nn}{\nonumber}
\ncmd{\mbf}[1]{\bs{#1}}
\ncmd{\gam}{\gamma}
\ncmd{\sig}{\sigma}
\ncmd{\pha}{\alpha}
\ncmd{\lam}{\lambda}
\ncmd{\dl}{\delta}
\ncmd{\kap}{\kappa}
\ncmd{\Lam}{\Lambda}
\ncmd{\Gam}{\Gamma}
\ncmd{\Dl}{\Delta}
\ncmd{\Ups}{\Upsilon}
\ncmd{\Om}{\Omega}
\ncmd{\eps}{\epsilon}
\ncmd{\veps}{\varepsilon}
\ncmd{\vphi}{\varphi}
\ncmd{\vtheta}{\vartheta}
\ncmd{\tw}{\text{w}}
\ncmd{\pd}{\partial}
\ncmd{\pll}{\parallel}
\ncmd{\mc}{\mathcal}
\ncmd{\mf}{\mathfrak}
\ncmd{\bs}{\boldsymbol}
\ncmd{\half}{\frac{1}{2}}
\ncmd{\tilJ}{\tilde{J}}
\ncmd{\avg}[1]{\langle{#1}\rangle}
\ncmd{\note}[1]{{\color{red}{\ding{168} [#1]}}}
\ncmd{\eq}[1]{Eq. \eqref{#1}}
\ncmd{\fig}[1]{Fig. \ref{#1}}
\ncmd{\suppl}{\note{`Supplementary Information'}}
\ncmd{\pg}[1]{\textcolor{red}{#1}}

\ncmd{\cl}[1]{CL$_{#1}$}
\ncmd{\Q}[1]{4Q$_{#1}$}
\ncmd{\an}[1]{{\color{blue} #1}}
\ncmd{\anc}[1]{\an{\footnotesize (AN) #1}}
\ncmd{\new}[1]{{\color{purple} #1}}

\begin{document}

\title{Field induced non-BEC transitions in frustrated magnets}

\author{Shouvik Sur$^1$, Yi Xu$^1$, Shuyi Li$^1$, Shou-Shu Gong$^2$, and Andriy H. Nevidomskyy$^1$}
\affiliation{$^1$Department of Physics and Astronomy, Rice University, Houston, TX 77005, USA}
\affiliation{$^2$School of Physical Sciences, Great Bay University, Dongguan 523000, China, and \\
Great Bay Institute for Advanced Study, Dongguan 523000, China}

\begin{abstract}
Frustrated spin-systems have traditionally proven challenging to understand, owing to a  scarcity of controlled methods for their analyses. 
By contrast, under strong magnetic fields, certain aspects of spin systems admit simpler and universal description in terms of hardcore bosons. 
The bosonic formalism is anchored by the phenomenon of Bose-Einstein condensation (BEC), which has helped explain the behaviors of a wide range of magnetic compounds under applied magnetic fields.
Here, we focus on the interplay between frustration and externally applied magnetic field to identify instances where the BEC paradigm is no longer applicable.
As a representative example, we consider the antiferromagnetic $J_1 - J_2 - J_3$ model on the square lattice in the presence of a uniform external magnetic field, and demonstrate that the frustration-driven suppression of  the  N\'{e}el order leads to a Lifshitz transition for the hardcore bosons. 
In the vicinity of the Lifshitz point, the physics becomes unmoored from the BEC paradigm, and the behavior of the system, both at and below the saturation field, is controlled by a Lifshitz multicritical point. 
We obtain the resultant  universal scaling behaviors, and provide strong evidence for the existence of a frustration and magnetic-field driven correlated bosonic liquid state along the entire phase boundary separating the N\'{e}el phase from other magnetically ordered states.
\end{abstract}

\date{\today}

\maketitle

\paragraph*{{\bf Introduction:}}
Bose-Einstein condensates and superfluids are the most generic ground states of repulsively-interacting, dense Bose gases above one dimension~\cite{leggett2001}.
For bosons hopping on a lattice, additional possibilities, such as Mott insulating phases, become possible at strong repulsive interactions~\cite{fisher1989}.
It has been suggested that, under suitable conditions,  interacting bosons may also exist in a symmetric quantum-liquid state -- a Bose metal, which is stabilized by an interplay between interactions and an enhanced low-energy density of states (DOS)~\cite{paramekanti2002, sur2019}. 
Over the past decades, the latter property has been utilized for stabilizing other kinds of Bose liquid states in Rashba spin-orbit coupled bosons~\cite{po2015}, deconfined critical points between valence bond solids~\cite{vishwanath2004}, superfluid phases in dipolar Bose-Hubbard model~\cite{lake2022dipolar}, certain tensor gauge theories~\cite{ma2018}, and fractonic superfluids~\cite{yuan2020}.
Unlike their fermionic counterparts,  pure bosonic systems are comparatively rare in nature.
It is, therefore, important to identify new platforms which may support unconventional phenomenology of  bosonic systems.

Due to the connection between localized spins and bosons, frustrated magnets are promising candidates for realizing  unconventional bosonic matter.
Frustrated magnetic systems, however, pose significant challenges to a theorist, owing to a scarcity of controlled approaches, especially for low-spin systems~\cite{balents2010, starykh2015}. 
A rare avenue becomes available in the presence of a uniform magnetic field -- since all spins in any quantum  magnetic system will polarize when exposed to a sufficiently strong magnetic field, quantum fluctuations are suppressed in the vicinity of the resultant  field-polarized (FP) state. 
In this region, the system can be mapped to a dilute gas of interacting bosons~\cite{beliaev1958}, and frustration manifests itself in the bosonic  band structure.
Indeed, much of the conventional phenomenology of interacting dilute Bose gases has been  realized in such magnetic systems, including BEC, superfluidity, and Mott transition~\cite{giamarchi2008, zapf2014}. 
Since the degree of frustration acts as an additional non-thermal tuning parameter, it introduces the possibility of realizing  unconventional states of bosonic matter~\cite{kamiya2014,wang2015,balents2016, jiang2022}, which bear similarities with those proposed in spin-orbit coupled bosonic  systems~\cite{galitski2013, zhai2015}. 
In this letter, we focus on the vicinity of  multicritical points that arise at the intersections of frustration-driven and magnetic-field-driven continuous phase transition lines. 
While frustration tends to stabilize  quantum paramagnetic states, a high magnetic field nearly saturates the spins.
As we shall show, the combined effect of the two non-thermal agents facilitates a  controlled  access to Bose liquid states in frustrated magnets under an applied magnetic field, which are analogs of Bose metals and have remained  unexplored in this context.

\begin{figure}[!t]
\centering
\subfloat[\label{fig:J2-J3}]{%
  \includegraphics[width=0.495\columnwidth]{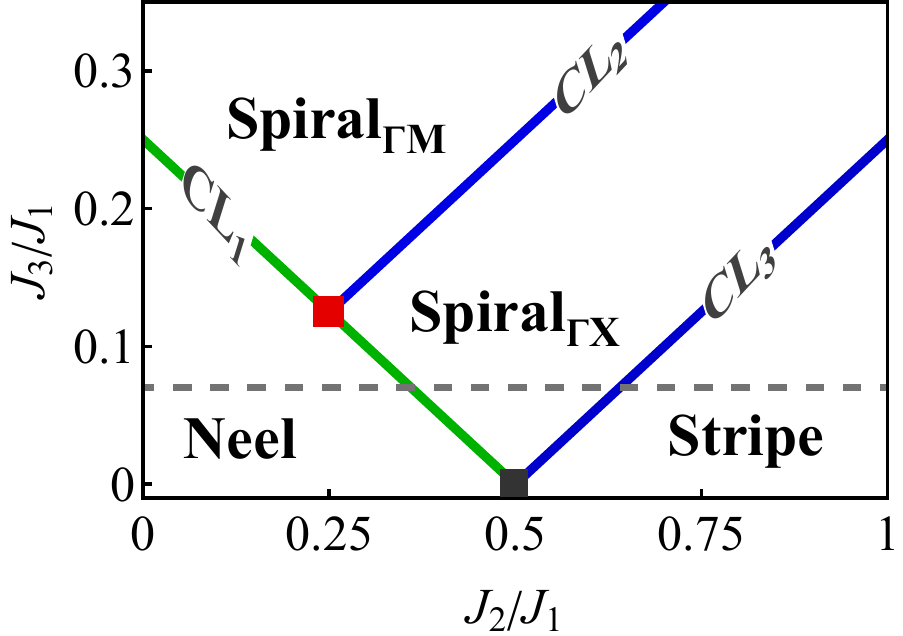}%
}
\hfill
\subfloat[\label{fig:h-J2}]{%
  \includegraphics[width=0.495\columnwidth]{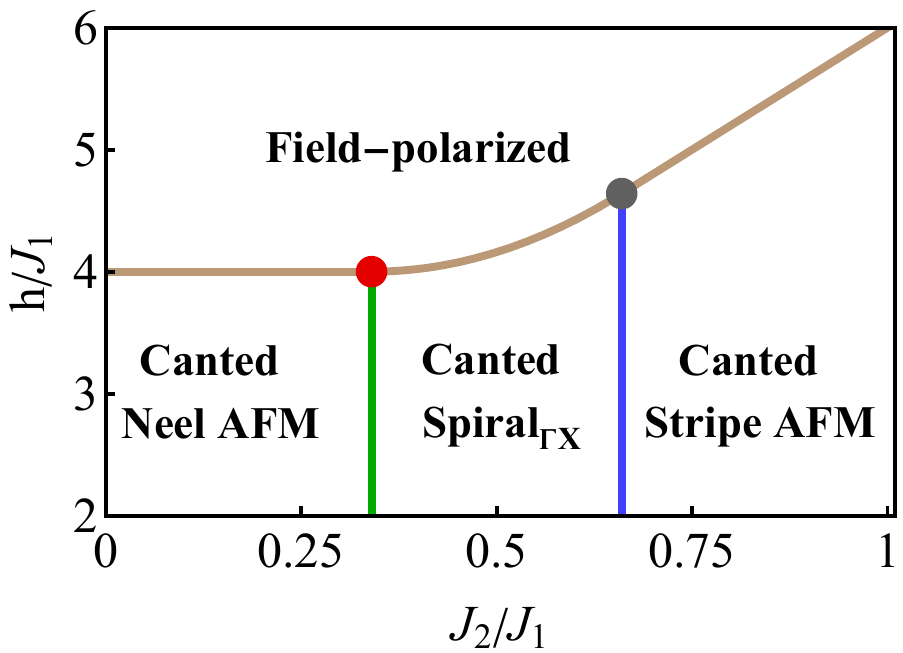}%
}
\hfill
\caption{
Phase diagrams in the absence and presence of an externally applied magnetic field ($h$).
(a) Classically, at $h =0$, four  antiferromagnetic phases are obtained, which are separated by critical lines (CL$_n$). 
(b) These phases develop canting with $h$, before continuously transitioning  to field-polarized states at sufficient high $h > h_c$ (brown curve). 
Multicritical points (filled squares and circles) are obtained at the intersection of all critical lines.
The phase boundaries in (b) are obtained from a linear spin-wave analysis at a fixed $J_3/J_1$ [dashed line in (a)].
}
\label{fig:phases}
\end{figure}

The zero-temperature transition between an FP and a magnetically ordered state is expected to be continuous, whereby the spin-rotational  symmetry perpendicular to the field-polarization direction is spontaneously broken.
The transition belongs to the `BEC universality class', which is characterized by the dynamical critical exponent $z = 2$~\cite{sachdev1994}.
Extensive experiments on antiferromagnets and  quantum paramagnets have established the importance of BEC-based perspective in understanding the physics of a wide variety of magnetic compounds under applied magnetic fields~\cite{ tanaka2001, cavadini2002, coldea2002, jaime2004,  paduan2004, waki2005, nakajima2006, garlea2007,  kitada2007, stone2007,  garlea2009,   aczel2009, tsirlin2009, thielemann2009,  samulon2009, bostrem2009, hong2010, tsirlin2011}. 
In this letter, we propose scenarios where this conventional outcome  breaks down. 
In particular, we establish (i) transitions that go beyond the BEC universality class, and (ii) explore the possibility of emergent Bose metallic physics in spin systems exposed to strong magnetic fields.
{We expect our results to be relevant to frustrated magnets with signatures of spin-liquid correlations  under high magnetic fields~\cite{qsl-exp-trgl-lat-NaYbO2,qsl-exp-trgl-lat-NaYbSe2,co-based-exp-bcao1,co-based-exp-bcao2}.}


\paragraph*{{\bf Model and Phase diagram:}}
We consider a spin-$\half$ Heisenberg model on the square lattice with {antiferromagnetic interactions beyond nearest-neighbor},
\begin{align}
& H_0 = J_1 \sum_{\langle \bs{r} \bs{r}' \rangle} \vec{S}_{\bs{r}} \cdot \vec{S}_{\bs{r}'} 
+ J_2 \sum_{\langle \langle \bs{r} \bs{r}' \rangle \rangle} \vec{S}_{\bs{r}} \cdot \vec{S}_{\bs{r}'}
+ J_3 \sum_{\langle \langle \langle \bs{r} \bs{r}' \rangle \rangle \rangle} \vec{S}_{\bs{r}} \cdot \vec{S}_{\bs{r}'} 
\label{eq:h0}
\end{align}
where all $J_n > 0$,   and $\vec S_{\bs r}$ represents the three-component spin-$1/2$ operator at site $\bs r$.
We employ $J_1$ as the overall energy scale, and define dimensionless ratios $\tilde X = X/J_1$ for any quantity $X$ that possesses the dimension of energy.
The classical phase diagram, obtained by analyzing Luttinger-Tisza (LT) bands~\cite{sm}, is presented in \fig{fig:J2-J3}. 
For $\tilJ_2 + 2\tilJ_3 < 1/2$ a N\'{e}el antiferromagnet (AFM) is realized.
In the complement of this region, classically, various spiral and stripe ordered phases are expected.
{The transitions between  N\'{e}el and spiral ordered phases are expected to be 2$^\text{nd}$ order, with a  continuous evolution of the ordering wavevector (see e.g. Ref.~\onlinecite{butcher2023}), which manifest themselves as Lifshitz transitions of the LT bandstructure.
The corresponding critical points  lie along the line $\tilJ_2 + 2\tilJ_3 = 1/2$ with $\tilJ_3>0$
}, henceforth labeled as `critical line 1' (\cl{1})~\cite{note-CLs}.
Because of the enhanced DOS on \cl{1}, quantum fluctuations may be expected to suppress magnetic order in its  vicinity~\cite{capriotti2004, kharkov2018, reuther2011, sindzingre2010}.
Recent numerical simulations support this expectation, and quantum spin-liquid states have been reported in the vicinity of \cl{1}~\cite{gong2014,liu2021, wu2022}.

We introduce a uniform magnetic field, $B$, such that the system is governed by 
$H(h) = H_0 - h  \sum_{\bs r} S_{\bs r}^{(z)}$,
where $h \coloneqq \mathfrak{g} \mu_B B$ is the 
Zeeman field
with $\mathfrak{g}$ and  $\mu_B$ denoting the Land\'{e} $g$-factor and Bohr magneton,  respectively.
The magnetic field tends to polarize the spins along $\hat z$ direction, and cants the AFM order.
At sufficiently high fields ($h > h_c$ with $h_c$ being the saturation field), the canted AFM phases give way to field-polarized (FP) states, which are classical ground states with all spins polarized along the magnetic field direction ($\hat z$).
A constant-$\tilJ_3$ slice of the  resultant phase diagram in the large-$S$ limit is depicted in Fig.~\ref{fig:h-J2}.
In this letter, we focus on the neighborhood of the transition between the canted AFM and FP phases.
In particular, we ask how the transition is affected by the Lifshitz criticality along \cl{1}.
We formulate a scaling theory for the  multi-critical points at the intersection of the saturation-field surface and \cl{1} (see Fig.~\ref{fig:h-J2}), and demonstrate the existence of magnetic field-tuned transitions belonging to a non-BEC universality class for \textit{all points on} \cl{1}.
These non-BEC critical points strongly affect 
the phase diagram in their vicinity, most remarkably through the stabilization of a quantum-liquid state 
at sub-critical fields.


\paragraph*{{\bf Non-BEC transitions:}}
In the vicinity of $h_c$, spin fluctuations may be conveniently modeled by density and phase fluctuations of  hardcore bosons  through the   Matsubara-Matsuda transformation~\cite{matsubara1956, batyev1984}, $S_{\bs{r}}^{(+)} \to b_{\bs{r}}^{\dagger}$; 
$S_{\bs{r}}^{(-)} \to b_{\bs{r}}$; 
$S_{\bs{r}}^{(z)} \to \half - \rho_{\bs{r}}$.
Thus, we rephrase the problem in terms of the hardcore bosons, $b_{\bs r}$, with  $\rho_{\bs r}$ being their local density.
The Hamiltonian acquires the form of a Bose-Hubbard model on the square lattice
\begin{align}
H(h) &= \int \frac{d^2 K}{(2\pi)^2}  \qty[\mc{E}(\bs K) - \mu(h)] b(\bs K)^\dagger b(\bs K) 
\nn \\
& + \int \frac{d^2 Q}{(2\pi)^2} V(\bs Q) \rho(-\bs Q) \rho(\bs Q) + U \sum_{\bs r} n_{\bs r}  (n_{\bs r} - 1),
\label{eq:H}
\end{align}
where the last term enforces the hardcore condition in the limit $U \to \infty$~\cite{batyev1984}.
The ``chemical potential'', $\mu(h) = \sum_{i=1}^3 J_i - h$, tuned by $h$,  controls the average density of bosons.
The dispersion, $\mc{E}(\bs K)$, and the  coupling function, $V(\bs Q)$, are independent of $h$, but sensitive to the $J_n$'s~\cite{sm}.
In particular, $\mc{E}(\bs K)$ tracks the LT band structure, and reflects the singularities at the classical phase boundaries: at a fixed $\tilJ_3$ and as a function of $\tilJ_2$, the boson band undergoes  Lifshitz transitions as the critical lines are crossed~\cite{note-Lifshitz}
We note that XXZ anisotropies, if present, can be absorbed in $V(\bs Q)$.

In the N\'{e}el AFM phase the dispersion is minimized at the $M$-point of the BZ.
Thus, the long-wavelength fluctuations of the bosons, $\Phi$,  carry momenta in the vicinity of the $M$-point, and the low-energy effective theory governing these fluctuations is given by $S_{M} = \int \dd{\tau} \dd{\bs r}  \mathcal{L}_{M}[\Phi(\tau, \bs r)]$ with  
\begin{align}
\mathcal{L}_{M}[\Phi] = \Phi^* \qty[\partial_\tau + \veps(\grad) - \mu_{\text{eff}}]\Phi + g |\Phi|^4,
\label{eq:LM}
\end{align}
where we have expanded the dispersion as $\mc{E}((\pi, \pi) + \bs k) = - \mc{E}_0 + J_1 \veps(\bs k)$ such that $\veps(\bs k) \geq 0$, and defined the effective parameters  $\mu_{\text{eff}} = J_1(\tilde h_c -  \tilde h)$ with $ \tilde h_c = (3 - \tilJ_2 - \tilJ_3)$, and $g \coloneqq \tilde V(\bs Q = \bs 0) = 2(1 + \tilJ_2 + \tilJ_3)$.
The magnetic field-driven transition can be understood as a transition between a state with no bosons (an FP state; $\mu_{\text{eff}} <0 \equiv h > h_c$) to a state with a finite density of bosons ($\mu_{\text{eff}} > 0 \equiv h < h_c $).
The transition itself is described with respect to the critical point at $\mu_{\text{eff}} = 0 \equiv h = h_c $.
If a magnetic long-range order is present for $h < h_c$, {the bosons develop an off-diagonal long-range order (ODLRO), which implies a BEC state~\cite{leggett2001,penrose1956}} with $\avg{\Phi} \neq 0$.
As CL$_1$ is approached from the N\'{e}el AFM side of the phase diagram, does the field-driven transition continue to be described by the BEC universality class?

\begin{figure}[!t]
\centering
\subfloat[\label{fig:cone}]{%
  \includegraphics[width=0.85\columnwidth]{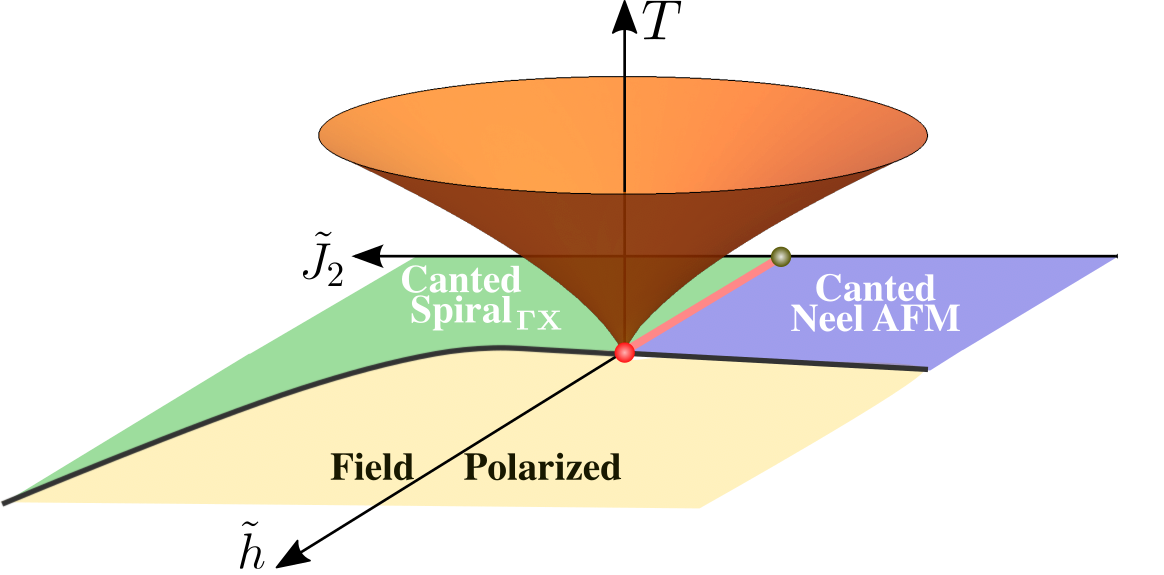}%
}
\hfill
\subfloat[\label{fig:cross}]{%
  \includegraphics[width=0.75\columnwidth]{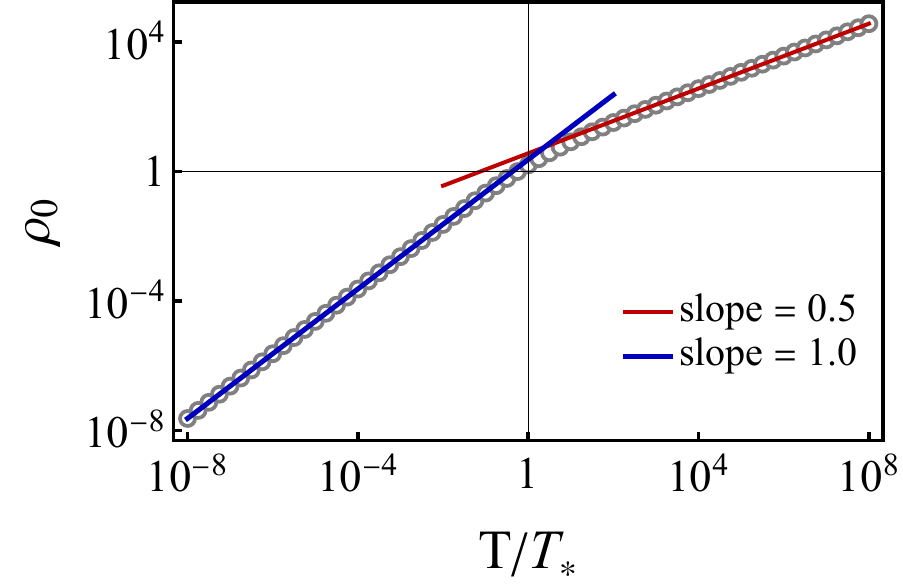}%
}
\caption{ Signatures of Lifshitz multicriticality. 
(a) The multicritical point (red dot) controls finite-$T$  behaviors of the system within  the (orange) critical-cone.
(b) Crossover behavior of $\rho_0$ with $T$ [c.f. Eq.~\eqref{eq:rho-T}].
The circles (lines) are numerically evaluated values of $\rho_0$ (fits to the data).
 The unequal slopes indicate a crossover from $\rho_0 \sim T \to \sqrt{T}$.
Here, $T_*$ is the temperature scale associated with the cone in (a).
}
\label{fig:mcp}
\end{figure}

We answer this question by first noting the dispersion about the band-minimum in the vicinity of \cl{1}, 
\begin{align}
\veps(\bs k, m_L) =
m_L|\bs k|^2 +  A \cos{\gamma} (k_x^4 + k_y^4) +  2 A \sin{\gamma} k_x^2 k_y^2,
\end{align}
where the `Lifshitz mass' $m_L = (1/2  - \tilJ_2 - 2\tilJ_3)$, and the parameters $A = \frac{1}{24} \sqrt{36 \tilJ_2^2 + (2 \tilJ_2 + 16 \tilJ_3 - 1)^2}$ and $\gamma = \tan^{-1}\qty{6\tilJ_2/(2\tilJ_2 + 16\tilJ_3 - 1)}$~\cite{sm}.
In the parameter regime where $m_L >0$, the field-driven transition belongs to the BEC universality class.
As \cl{1} is approached, $m_L \to 0$ and the field-driven transition belongs to a distinct universality class that is controlled by the Lifshitz multi-critical point (LMCP) at $h = h_c$ and $\tilJ_2 = \tilJ_{2c}$.
At the LMCP, although  $\mu_{\text{eff}} = 0$,  strong quantum fluctuations arise in the presence of interactions among bosons, owing to the divergent DOS.
Consequently,  $V_{\text{eff}}$ becomes  strongly relevant at the Gaussian fixed point governed by the first term in \eq{eq:LM}.
This strong coupling theory, however, is exactly solvable at $T=0$, due to the absence of particle-hole excitations~\cite{beliaev1958,sachdev1994, sachdevBook}.
In particular, the positive semi-definiteness of $\veps(\bs q)$ leads to a chirality-like constraint on the  bosonic-dynamics, which protects the quadratic terms in the action against quantum corrections~\cite{sachdev1994b, sm}.
This is analogous to chiral fermionic liquids, where tree-level or classical critical exponents remain robust against quantum fluctuations, thanks to the chiral dynamics~\cite{wen1990, sur2014}.
Thus, in the present case, the tree-level critical exponents,
\begin{align}
z = 4; \quad \nu_h = 1/4; \quad \nu_J = 1/2; \quad \eta = 0,
\end{align}
do not accrue anomalous dimensions  through quantum fluctuations~\cite{sm}.
Here, $z$ is the dynamical critical exponent, $\nu_h$ and $\nu_J$ control the scaling of the correlation length along $h$ and $J_2$ axes, respectively, and $\eta$ is the anomalous dimension of $\Phi$.
Since this is a multi-critical point, the correlation length with respect to the LMCP is given by $\xi =1/\sqrt{\xi_h^{-2}+\xi_J^{-2}} $ 
with $\xi_h \sim |h - h_c|^{-\nu_h}$ and $\xi_J \sim |J_2 - J_{2c}|^{-\nu_J}$. 
The critical exponents imply the magnetic field-driven transition at $J_2 = J_{2c}$ does not belong to the BEC universality class, which would have been  characterized by $\xi \sim |h - h_c|^{-1/2}$.

In contrast to the particle-hole channel, non-trivial quantum fluctuations are present in the particle-particle channel, which drive the system towards an interacting fixed point.
To see this, we perform Wilsonian renormalization group (RG) analysis at $d=4 -\epsilon$, where $d$ is the number of spatial dimensions.
We obtain the following one-loop RG flow of the parameters in $\mathcal L_M$~\cite{sm}:
\begin{align}
\partial_\ell \bar g = \eps \bar g - \frac{f_g(\gamma)~\bar g^2}{16 \pi^2 A} ,
\quad 
\partial_\ell \bar \mu = 4 \bar \mu,
\quad 
\partial_\ell \bar m_L = 2 \bar m_L,
\label{eq:rg}
\end{align}
where $\ell$ is the logarithmic length-scale, $\qty(\bar g, \bar \mu, \bar m_L) = \qty(\Lambda^{-\eps} g, \Lambda^{-4} \mu_{\text{eff}} , \Lambda^{-2} m_L )$, $\Lambda$ is the ultraviolet (UV) momentum cutoff, and  $f_g(\gamma) = \int_0^1  \dd{t}\{ [t^2+(1-t)^2] \cos{\gamma} +2 (1-t) t \sin{\gamma}\}^{-1}$.
Since the LMCP  is a multicritical point, it has two independent relevant directions, $\bar \mu$ and $\bar m_L$.
By maintaining multicriticality of the LMCP, i.e. setting the bare values $\bar m = 0 = \bar \mu$, we obtain a stable fixed point at  $(\bar g_*, \bar \mu_*, \bar m_{L, *}) = \qty(16\pi^2   A f_g^{-1}(\gamma) \eps,0 ,0)$.
Extrapolating the result to $\eps = 2$, yields a fixed-point coupling $\bar g_* = 32\pi^2  A f_g^{-1}(\gamma)$, which is independent of the UV structure of the interaction vertex, such as XXZ anisotropies.
{Because of its dependence on $A$ and $\gamma$, $\bar g_*$ varies along \cl{1}, as shown in Fig. S2 of~\cite{sm}.}
In particular, as the critical point at $(A, \gamma) = (\frac{1}{8}, \frac{\pi}{2}) \equiv (\tilJ_2, \tilJ_3) = (\frac{1}{2}, 0)$ is approached along \cl{1}, $f_g(\gamma) \sim \ln{[1/(\pi/2 - \gamma)]} \gg 1$; consequently, the fixed point is pushed to weaker  couplings, and the one-loop result   appears to become more accurate as $\gamma \to \pi/2$.

\paragraph*{{\bf Multicriticality and crossover behaviors:} } 
The LMCP is an example of `zero-scale-factor universality', and the scaling functions for all observables are completely determined by microscopic or bare  parameters~\cite{sachdev1994}.
Here, we focus on finite-temperature properties within the multi-critical cone emanating from the LMCP, as depicted in Fig.~\ref{fig:cone}.
The shape of the cone is controlled by the temperature scale, $T_* = \sqrt{T_{*, h}^2 + T_{*, J}^2}$ with  $T_{*, h} \sim  \xi_h^{-z} \sim |h - h_c|$ and $T_{*, J} \sim  \xi_J^{-z} \sim |J_2 - J_{2c}|^{2}$.
Although the density of bosons at $h= h_c(\tilJ_2)$ vanishes at $T=0$, thermal fluctuations at $T>0$ makes it finite. 
Therefore, we expect the magnetization at $T>0$ would be suppressed below that in the FP state.
Using a finite-$T$ scaling analysis~\cite{fisher1988, sachdev1994}, we estimate the average boson density to scale as 
\begin{align}
\rho_0(T) \equiv \avg{\rho(T)} = T^{d/4}~f_T(T_{*}/T), 
\label{eq:rho-T}
\end{align}
where the dimensionless function has the limiting behavior, $\lim_{x \ll 1} f_T(x) = \order{1}$ and $\lim_{x \gg 1} f_T(x) \sim 1/\sqrt{x}$.
At the LMCP $T_*$ vanishes, and only the former limit is applicable.
In $d=2$ this leads to  $\rho_0(T) \equiv [\frac{1}{2} - \avg{S_{\bs r}^{(z)}}] \sim \sqrt{T}$.
Away from the LMCP, but along the BEC-transition line, $\rho_0(T)$ displays a crossover behavior.
At low temperatures  ($T \ll T_{*}$) the BEC critical points dictate the scaling and $\rho_0 \sim T$.
At sufficiently high temperatures ($T \gg T_{*}$), however, the system enters the critical cone  and $\rho_0 \sim \sqrt{T}$.
This crossover behavior is depicted in Fig.~\ref{fig:cross}.

\begin{figure}[!t]
\centering
\includegraphics[width=0.9\columnwidth]{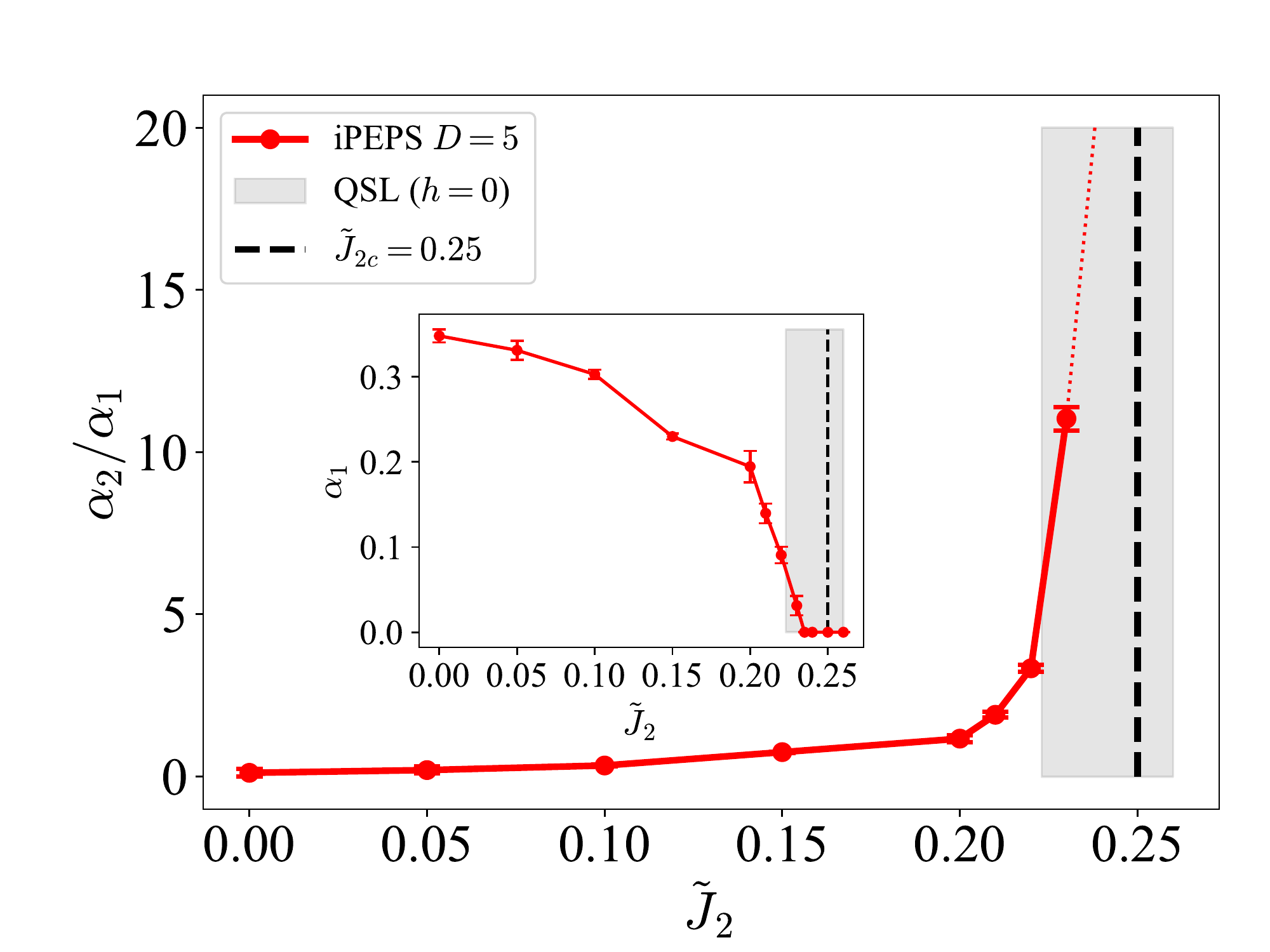}%
\caption{
Crossover in the scaling of $[1/2 - \avg{S^{(z)}}]$ with  $\Delta h \coloneqq (h_c - h)$ with  increased frustration, obtained from iPEPS simulations.
The data is fitted to the function,  $[1/2 - \avg{S^{(z)}}] = \alpha_1 \Delta h \ln \frac{h_c}{\Delta h} + \alpha_2 \sqrt{\Dl h}$.
Deep in the N\'{e}el phase the transition belongs to the Bose-Einstein-condensation  universality class; consequently,  $\alpha_1$ is $\order{1}$  (inset) and $\alpha_2 \ll 1$.
Upon approaching the classical phase boundary, the ratio  $\alpha_2/ \alpha_1$ increases with $\alpha_1 \to 0$.
The shaded region indicates the regime where a quantum spin liquid state has been reported at $h=0$~\cite{liu2021}.
The dotted (dashed) line is an extrapolation of the data towards $\tilJ_{2c}$ (marks $\tilJ_{2c}$).
Here, we have fixed $\tilJ_3 = 1/8$. 
}
\label{fig:peps}
\end{figure}


The LMCP's influence on the  phase diagram at sub-critical fields can be understood in terms of the density and phase fluctuations of the bosons.
While a finite mean-density reflects the deviation of $\avg{S^{(z)}}$ from $1/2$, phase fluctuations determine the correlation between $S^{(+)}$ and $S^{(-)}$.
First, we consider the asymptotic behavior of the mean density in the region $0<(1 - h/h_c) \ll 1$, which corresponds to $0<\mu_{\text{eff}} \ll J_1$.
From one-loop RG analysis, we obtain the scaling of the mean density with $\mu_{\text{eff}}$,
\begin{align}
\rho_0(\mu_{\text{eff}}) = \mu_{\text{eff}}^{d/4} ~f_h\qty(m_L^2/\mu_{\text{eff}}).
\end{align}
The dimensionless scaling function, $f_h(x)$, is such that $\lim_{x \ll 1} f_h(x) = \order{1}$ and $\lim_{x \gg 1} f_h(x) \sim 1/\sqrt{x}$.
Therefore, for a fixed $\mu_{\text{eff}}/J_1$ at $d=2$, as the system is tuned towards the LMCP from the canted N\'{e}el phase, the asymptotic scaling of $\rho_0 = [1/2 - \avg{S^{(z)}}]$  crosses over from $\rho_0 \sim (h_c - h) \to (h_c - h)^{1/2}$.
We verify this crossover behavior through unbiased numerical calculations using infinite projected entangled-pair states (iPEPS)~\cite{pepstorch} as demonstrated in Fig.~\ref{fig:peps}. 
{We note that iPEPS works directly in the thermodynamic limits by exploiting translation invariance~\cite{ipeps}. 
The accuracy of this variational ansatz is controlled by
the bond dimension $D$ of the tensors involved in their construction, which is related to the entanglement of the state.}

\paragraph*{{\bf Emergent algebraic liquid:}}
In order to understand the behavior of phase fluctuations at sub-critical fields we introduce the  hydrodynamic variables,  $\vtheta$ and $\varrho$, which represent the long-wavelength phase and density fluctuations, respectively, of boson field,
\begin{align}
\Phi(\tau, \bs r) = \sqrt{\rho_0 + \varrho(\tau, \bs r)} ~ e^{i \vtheta(\tau, \bs r)}.
\label{eq:hydro}
\end{align}
For $\tilJ_2 < \tilJ_{2c}$ the FP state transitions into a canted N\'eel-AFM as $h$ is lowered below $h_c$.
This phenomenon is reflected in an $U(1)$ symmetry breaking transition for the bosons, whereby $\avg{\Phi} \sim \sqrt{\rho_0} e^{- \frac{1}{2} \avg{\vtheta^2} } \neq 0$, which  implies existence of an off-diagonal long-range order (ODLRO), hence a BEC~\cite{penrose1956,leggett2001}. 
As $\tilJ_2 \to \tilJ_{2c}$, the condensate fraction $\sim \avg{\Phi}$ is suppressed due to increased phase fluctuations.
What is the fate of the system as $\avg{\Phi} \to 0$? 

\begin{figure}[!t]
\centering
\includegraphics[width=0.75\columnwidth]{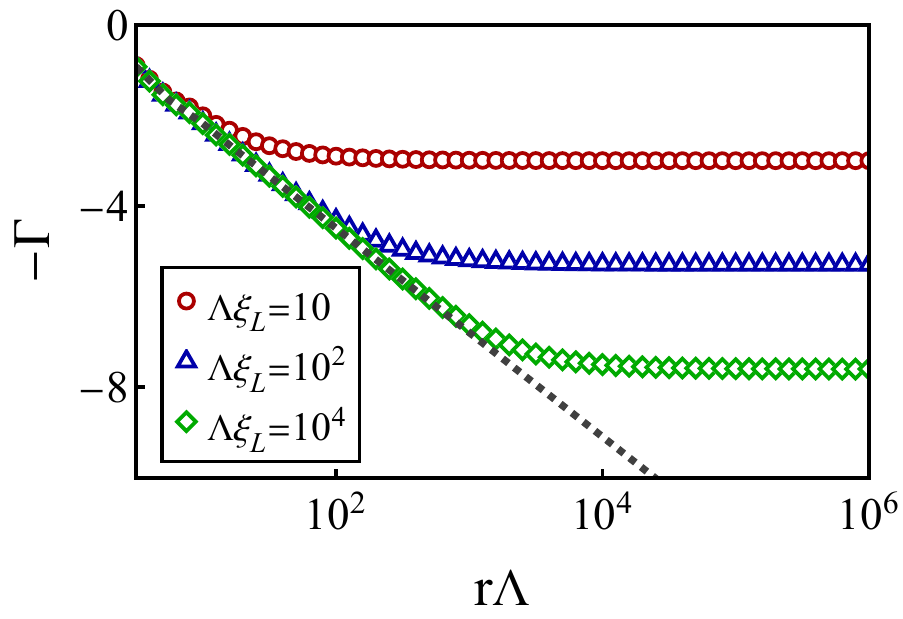}%
\hfill
\caption{Crossover behavior of $\langle S^{(+)}_{\bs 0} S^{(-)}_{\bs r} \rangle$ as a function of $\bs r$.
For $|\bs r| \Lambda \gg 1$ ($|\bs r| \Lambda \ll 1$) the behavior is controlled by the canted-N\'{e}el phase (quantum critical point at $\tilJ_2 = \tilJ_{2c}$). The dotted line represents algebraic decay per \eq{eq:PM}.
Here, $\xi_L \propto (\tilJ_{2c} - \tilJ_2)^{-1/2}$ and $\Lambda^{-1}$ is a short-distance cutoff.
}
\label{fig:crossover}
\end{figure}

The dynamics of $\Phi$, as dictated by $S_M$, is controlled by \textsl{two} independent length scales, $\rho_0^{-1/2}$ and $m_L^{-1/2}$.
We fix the mean density $\rho_0$ (for fields $h<h_c$) and consider the influence of $m_L$ (which controls proximity to \cl{1}) on the dynamics.
The phase fluctuations are governed by the effective action~\cite{sm} 
\begin{align}
& S_\vtheta = \int \frac{dk_0 d\bs k}{(2\pi)^3} \qty[\frac{k_0^2}{4g} + \rho_0 \veps(\bs k, m_L)] \vtheta(-k) \vtheta(k),
\label{eq:S-theta}
\end{align}
where $k_0$ is the Euclidean frequency.
We note that the propagator of $\vtheta$ is non-perturbative in $g$, and the phase fluctuations disperse as $\sqrt{4g\rho_0 \veps(\bs k, m_L)}$, which is analogous to the dispersion of magnons in the canted N\'{e}el phase.
The long-wavelength behavior of the equal-time correlation function,
\begin{align}
\langle S^{(+)}_{\bs 0} S^{(-)}_{\bs r} \rangle 
\sim \langle \Phi^\dagger(0, \bs 0) \Phi(0, \bs r) \rangle 
= \rho_0 \exp[- \Gamma(\bs r, \xi_L)],
\label{eq:Spm}
\end{align}
is determined by the correlation length $\xi_L \equiv \sqrt{A/m_L}$ through $\Gamma(\bs r, \xi_L)$~\cite{sm}.
The function $\Gamma(\bs r, \xi_L)$ is most easily computed along the line $\gamma = \pi/4$, on which $\veps(\bs k, m_L)$ acquires  an $C_\infty$-rotational  symmetry  and $\xi_L \sim \sqrt{\tilJ_2/(1-4\tilJ_2)}$.
As shown in Fig.~\ref{fig:crossover}, for $|\bs r| \gg \xi_L$, $\langle S^{(+)}_{\bs 0} S^{(-)}_{\bs r} \rangle$ saturates to a non-universal value (dependent on $\rho_0$ and $\xi_L$), implying the presence of ODLRO in $\Phi$~\cite{note-suppression}.
In the opposite limit, a universal scaling is obtained, indicating the presence of a quantum critical point (QCP) as $\xi_L \to \infty$ (dashed line in Fig.~\ref{fig:crossover}).
This putative QCP  is characterized by the absence of an BEC, i.e. $\avg{\Phi} = 0$.
At small but finite-$T$ the canted N\'{e}el phase possesses only a quasi-long-range order, and goes through a  Berezinskii-Kosterlitz-Thouless (BKT) transition upon raising $T$.
Since the BKT transition scale, $T_{\text{BKT}}$, is controlled by $m_L$, it is expected to be suppressed as \cl{1} is approached.
Thus, the resultant crossover behavior is  controlled by the critical fan emanating from the critical point at $m_L = 0 \equiv \tilJ_2 = \tilJ_{2c}$ for $h < h_c$ (see Fig.~\ref{fig:h-J2}).

Interestingly, the QCP realizes a higher-dimensional analog of the   Luttinger liquid, where a condensate cannot form due to strong infrared fluctuations.
{For sufficiently strong magnetic fields, and in the absence of proliferation of vortices of $\Phi$~\cite{note-vortex},
}
 all points on \cl{1} host such algebraic liquid states, which are parameterized by the critical exponent $\mc W$ that controls the long-wavelength behavior of transverse spin correlations:
\begin{align}
\langle S^{(+)}_{\bs 0} S^{(-)}_{\bs r} \rangle 
\sim \rho_0 (|\bs r| \Lambda)^{-\mc W}.
\label{eq:PM}
\end{align}
We find that  
$\mc W  = \sqrt{g/(\rho_0 A)}~f_{w}(\gamma)$, with $f_w$ being a dimensionless function~\cite{sm}.
While generic points on \cl{1} possess a $C_4$ rotational symmetry, an  $C_\infty$ symmetry emerges at $\gam = \pi/4$, where  \cl{1} and \cl{2} intersect (see Fig.~\ref{fig:phases}). 
The $C_\infty$ critical point would be expected to control the high-energy behavior in its vicinity, including that along \cl{2} where a different kind of higher-dimensional Luttinger liquid is expected~\cite{sur2019, lake2022}.

\paragraph*{{\bf Conclusion:}} 
Motivated by the ability of frustration to stabilize unconventional states of matter in quantum-spin systems, we studied its  interplay with an applied magnetic field.
With the help of the $J_1-J_2-J_3$ antiferromagnetic Heisenberg model, we demonstrated that frustration
limits the validity of the BEC paradigm in describing the approach to saturation field.
In particular, the phase transition between magnetically ordered and field-polarized states no longer belongs to the BEC universality class on the critical line  \cl{1},  along which frustration suppresses magnetic order.
A similar outcome is expected along \cl{2} and \cl{3}, where the corresponding transitions are governed by  distinct non-BEC universality classes.

In the vicinity of \cl{1},  at sub-critical fields, it is possible to realize bosonic quantum-liquid states that are stabilized by a combination of frustration and high magnetic fields.
These quantum-liquids are higher-dimensional analogs of gapless states that develop under sufficiently high magnetic fields in the spin-1 Haldane chain~\cite{sachdev1994} and 1D  valence bond solids~\cite{iaizzi2015}.
We note that mechanisms similar to that described here may be responsible for stabilizing the quantum spin-liquid phase in the Kitaev honeycomb compass  model in magnetic field along the [111] direction~\cite{patel2019}.
A detailed
investigation into such possibilities is left to future works.

\begin{acknowledgments}
The authors thank Andrey Chubukov, Oleg Tchernyshyov, and Nandini Trivedi for fruitful discussions. The analytical work performed by S.S. and A.H.N. was supported by the U.S. Department of Energy Computational Materials Sciences (CMS) program under Award Number DE-SC0020177. The numerical calculations performed by Y.X. were supported by the U.S. National Science Foundation Division of Materials Research under the Award DMR-1917511. The computing resources at Rice University were supported in part by the Big-Data Private-Cloud Research Cyberinfrastructure MRI-award funded by NSF under grant CNS-1338099 and by Rice University's Center for Research Computing (CRC). S.S.G. was supported by the NSFC (11834014, 11874078) and the Dongguan Key Laboratory of Artificial Intelligence Design for Advanced Materials. S.S. and A.H.N. are grateful for the hospitality of the Aspen Center for Physics, which is supported by National Science Foundation grant PHY-2210452.
\end{acknowledgments}



\appendix
\onecolumngrid

\section{Derivation of the effective theory}
\begin{figure}[h]
\centering
\subfloat[\label{fig:BZ}]{%
  \includegraphics[width=0.25\columnwidth]{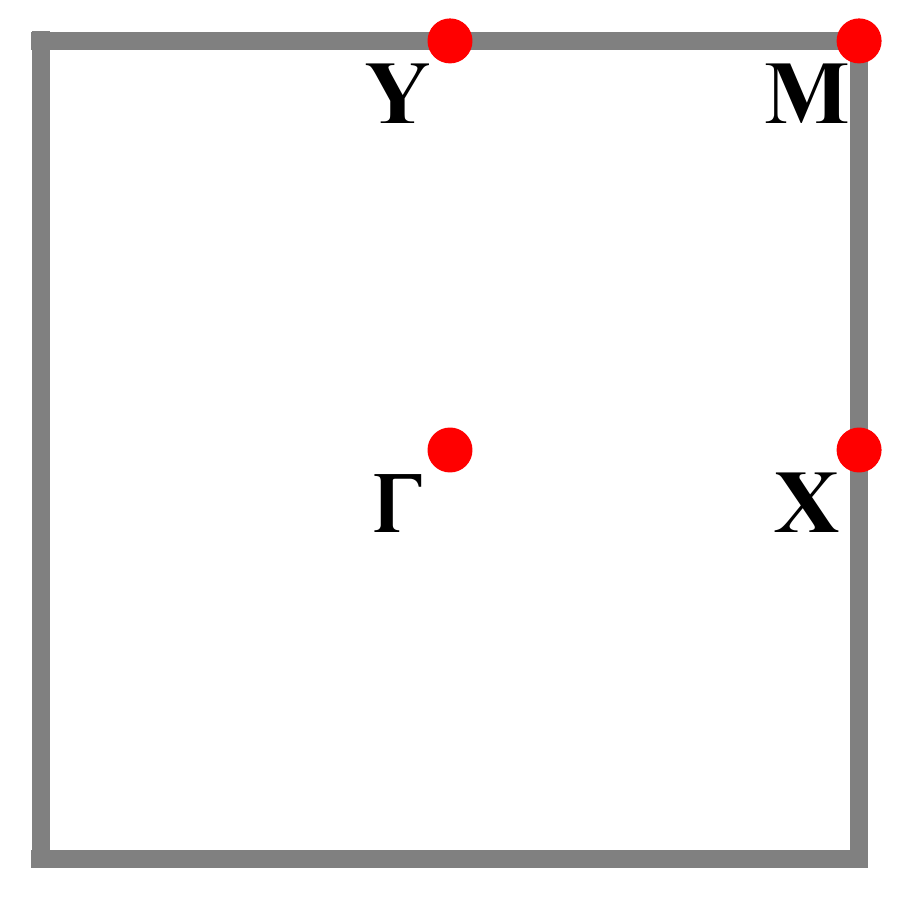}%
}
\hfill
\subfloat[\label{fig:1Q}]{%
  \includegraphics[width=0.3\columnwidth]{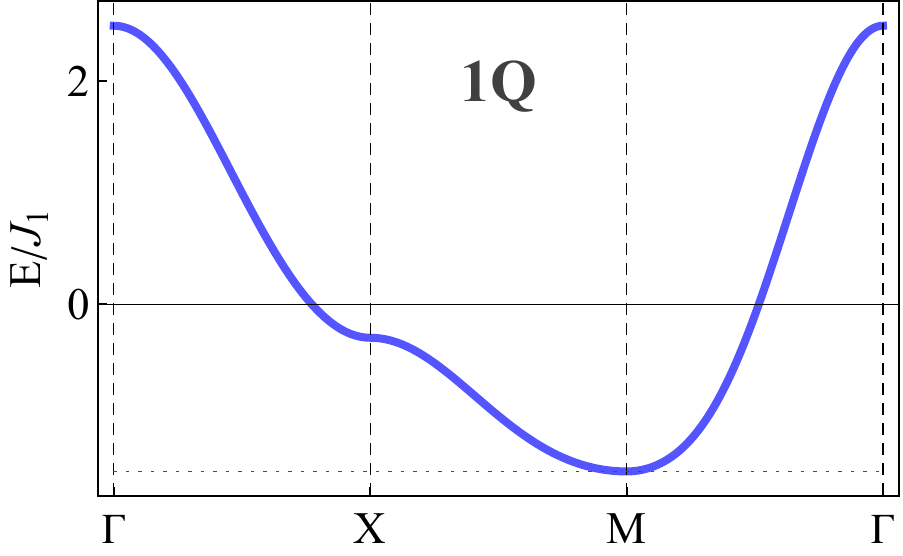}%
}
\hfill
\subfloat[\label{fig:2Q}]{%
  \includegraphics[width=0.3\columnwidth]{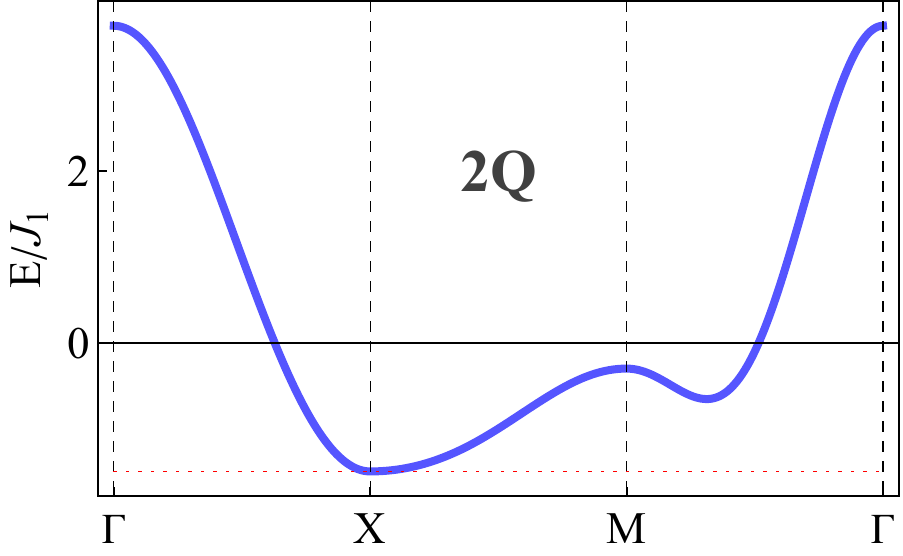}%
}
\hfill
\subfloat[\label{fig:4Q1}]{%
  \includegraphics[width=0.3\columnwidth]{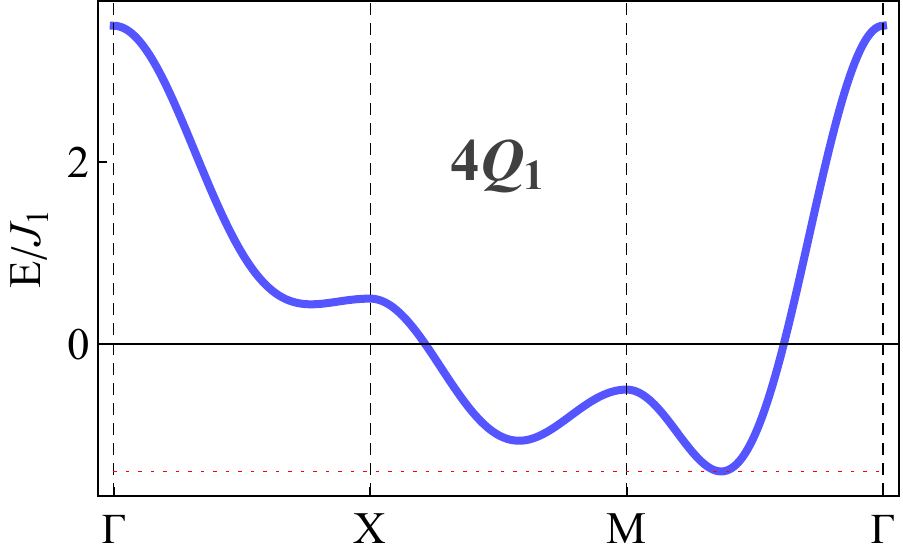}%
}
\hfill
\subfloat[\label{fig:4Q2}]{%
  \includegraphics[width=0.3\columnwidth]{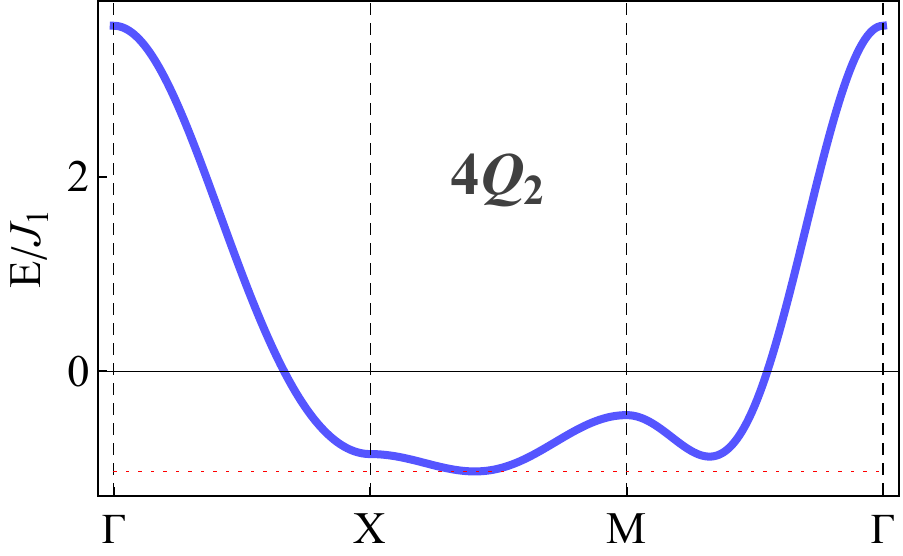}%
}
\hfill
\subfloat[\label{fig:cl1}]{%
  \includegraphics[width=0.3\columnwidth]{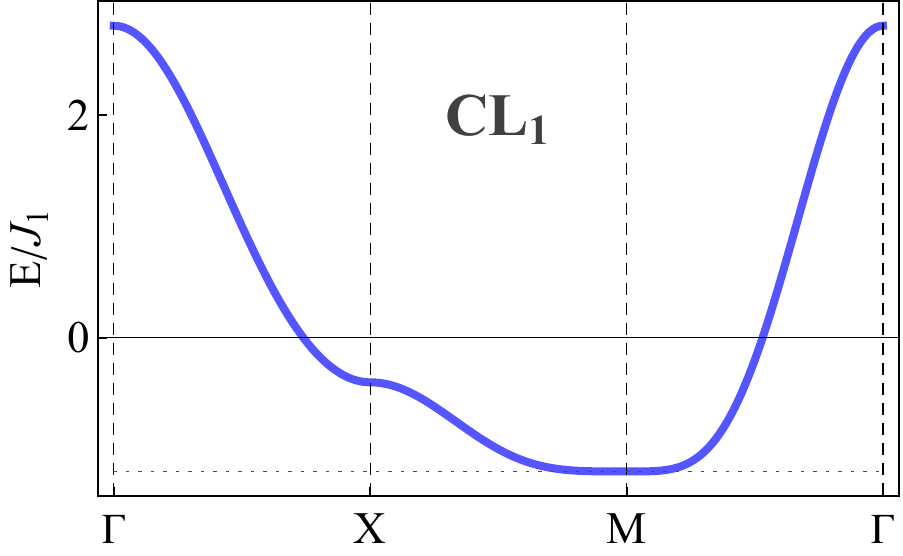}%
}
\hfill
\subfloat[\label{fig:cl2}]{%
  \includegraphics[width=0.3\columnwidth]{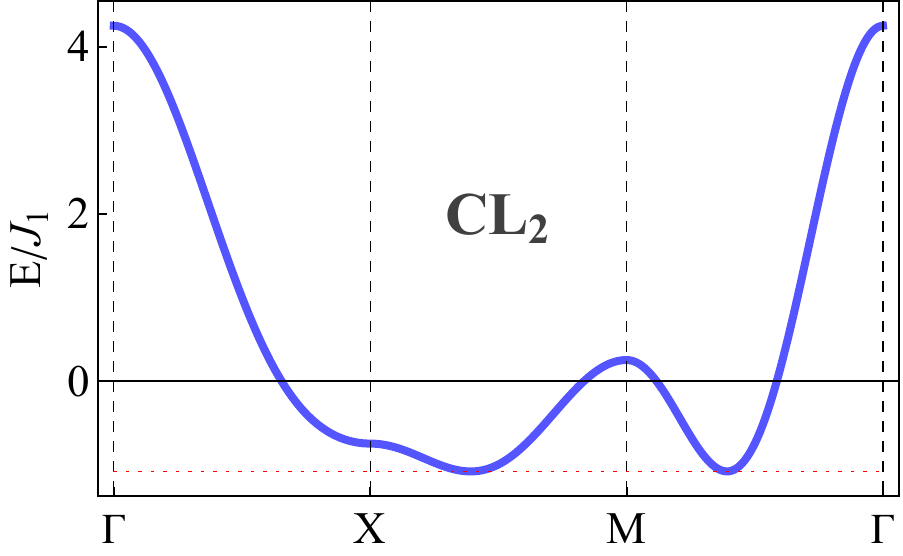}%
}
\hfill
\subfloat[\label{fig:cl3}]{%
  \includegraphics[width=0.3\columnwidth]{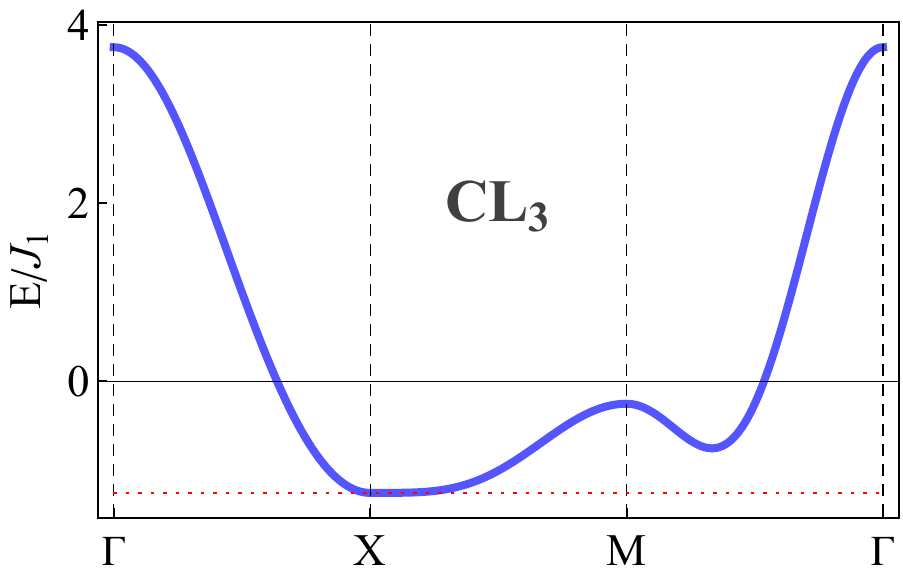}%
}
\hfill
\subfloat[\label{fig:mcp12}]{%
  \includegraphics[width=0.3\columnwidth]{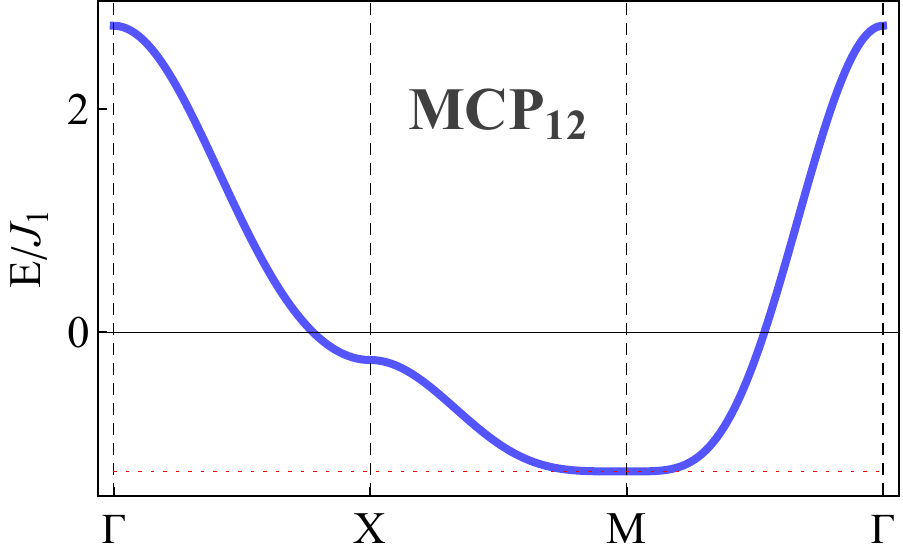}%
}
\hfill
\subfloat[\label{fig:mcp13}]{%
  \includegraphics[width=0.3\columnwidth]{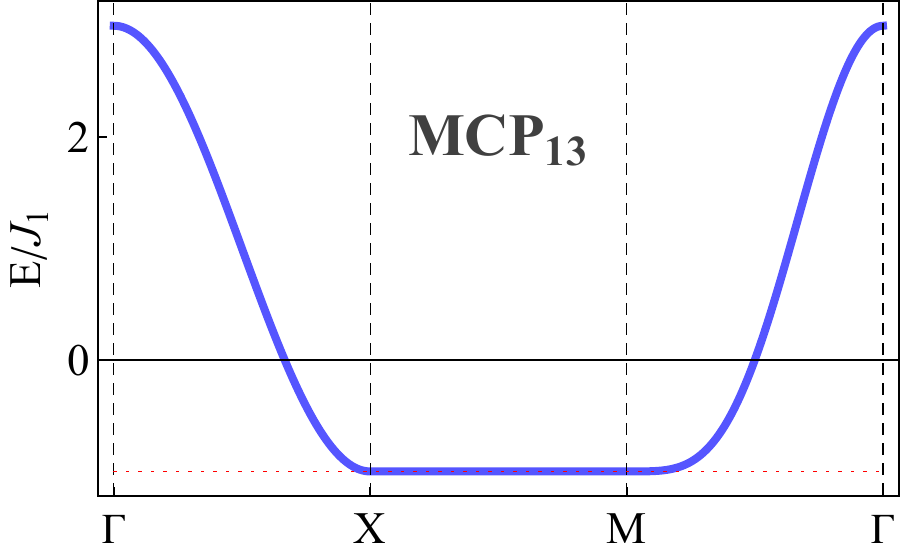}%
}
\caption{Detail of the dispersion  $\mc E(\bs K)$, in different phases in the phase diagram in Fig. 1 of the main text.
(a) The high symmetry points in the square Brillouin zone.
While in the (a) N\'{e}el, (b) stripe, and (d) \& (e) spiral ordered phases, respectively, $\mc E(\bs K)$ achieves one, two, and four minima. The labels on the plots refer to the number of independent ordering vectors supported by the corresponding phases.
Since in terms of $\mc E(\bs K)$ the transition between any pair of neighboring phases involves a reconstruction of the band structure, these can be classified as Lifshitz transition for the bosons; consequently, the low-energy density of states  acquires a singular form on the critical lines, viz. (f) $\rho(\veps) \sim \veps^{-1/2}$, (g) $\rho(\veps) \sim \veps^{-1/2}$, and (h) $\rho(\veps) \sim \veps^{-1/4}$.
At the intersection of a pair of critical lines multicritical points arise, (i) and (j).
Here, the notation MCP$_{ij}$ indicates intersection between \cl{i} and \cl{j}.
}
\label{fig:LT}
\end{figure}

In this section we derive the effective field theory in Eqs. (3) and (4) of the main text.
After Fourier transforming $b_{\bs r} = \int \frac{\dd{K_x} \dd{K_y}}{(2\pi)^2} e^{i \bs K \cdot \bs r} b(\bs K)$, we obtain the Bose--Hubbard Hamiltonian in the form of Eq.~(2) in the main text: 
\begin{align}
H(h) = \int \frac{d^2 K}{(2\pi)^2}  \qty[\mc{E}(\bs K) - \mu(h)] b(\bs K)^\dagger b(\bs K) 
 + \int \frac{d^2 Q}{(2\pi)^2} V(\bs Q) \rho(-\bs Q) \rho(\bs Q) + U \sum_{\bs r} n_{\bs r}  (n_{\bs r} - 1),
\label{eq:H_append}
\end{align}
where
\begin{align}
& \mc E(\bs K) = J_1 [\cos{K_x} + \cos{K_y}] + 2 J_2 \cos{K_x}\cos{K_y} + J_3 [\cos{(2K_x)} + \cos{(2K_y)}]; \\
& V(\bs Q) = \mc E(\bs Q); \\
& \mu = \sum_{n=1}^3 J_n - h.
\end{align}
A detailed analysis of the behavior of $\mc E(\bs K)$ across the phase diagram is presented in Fig.~\ref{fig:LT}.

In the parameter regime for the N\'{e}el ordered phase, $\mc E(\bs K)$ is minimized at $\bs K = (\pi, \pi)$.
Therefore, we expand around $\bs K = (\pi, \pi)$ to obtain the low energy dispersion for the bosons,
\begin{align}
J_1 \veps(\bs k) = \mc E((\pi, \pi)+\bs k) - \mc E((\pi, \pi)) = J_1 \qty[\underbrace{\frac{1 - 2 \tilJ_2 - 4\tilJ_3}{2}}_{m_L} |\bs k|^2 + \underbrace{\frac{16\tilJ_3 + 2\tilJ_2 -1}{24}}_{A \cos\gamma}(k_x^4 + k_y^4) + 2 \underbrace{\frac{\tilJ_2}{4}}_{A \sin\gamma} k_x^2 k_y^2 + \order{|\bs k|^6} 
],
\end{align}
where $\mc E((\pi, \pi)) = 2(J_2 + J_3 - J_1)$.
The effective chemical potential that appears in Eq.~(3) in the main text and in Eq.~\eqref{eq:effS} is defined as 
\begin{align}
\mu_{\text{eff}} = \mu - \mc E((\pi, \pi)) = \underbrace{J_1(3 - \tilJ_2 - \tilJ_3)}_{h_c} - h.
\end{align}
For $m_L \neq 0$, $\veps(\bs k) \sim |\bs k|^2$ and the density of state $\rho(\veps) \sim \veps^0$.
At $m_L = 0 $, however, $\veps(\bs k) \sim |\bs k|^4$ and $\rho(\veps) \sim \veps^{-1/2}$.
The specific energy-scaling of $\rho(\veps)$ is reminiscent of 1D bosons.

At low energies, since the only scattering channels available are those that involve momentum eigenstates in the vicinity of  $\bs K = (\pi, \pi)$, $\bs Q \approx 0$ and the most relevant short-range interaction channel is given by $V(\bs Q = 0)$.
We note that this is not the case in the stripe ordered phase where $\mc E$ is minimized at two inequivalent, but symmetry-related high-symmetry points, $\bs K = (0, \pi)$ and $(\pi, 0)$. 
In this case an additional scattering channel is available that mix bosonic modes carrying momenta in the vicnity of $\bs K = (0, \pi)$ with those near $\bs K = (\pi, 0)$; consequently, two scattering channels are important, $V(0)$ and $V(\pi, \pi)$.
Therefore, the approach to the critical line $m_L = 0$ is simplest from the N\'{e}el side of the phase diagram.

By fixing $1/4 > \tilJ_3 > 0$, it is clear that $m_L > 0$ [$m_L < 0$] for $\tilJ_2 < \half(1-4\tilJ_3)$ [$\tilJ_2 > \half(1-4\tilJ_3)$].
Thus, the transition between the N\'{e}el and spiral ordered phase is mapped to a transition driven by $m_L$, whereby the number of minima of the effective dispersion, $\veps(\bs k)$, changes from $1$ ($m_L > 0$) to $2$ ($m_L < 0$).
This is an example of a Lifshitz transition.
The critical point, $m_L = 0$, is the definition of \cl{1}.

\section{Scaling theory} \label{sec:rg}
Here, we derive the scaling relations and present our renormalization group (RG) analyses.
We begin with the effective field theory,
\begin{align}
S_{M} = \int \dd{k} \Phi^*(k) \qty[ik_0 + \veps(\bs k) - \mu_{\text{eff}}] \Phi(k) + g \int \dd{k_1} \dd{k_2} \dd{q} \Phi^*(k_1+q)\Phi(k_1) \Phi^*(k_2)\Phi(k_2+q).
\label{eq:effS}
\end{align}
Thus, the propagator of $\Phi$ is given by
\begin{align}
G_0(k) = \frac{1}{ik_0 + \veps(\bs k) - \mu_{\text{eff}}}.
\end{align}
As mentioned in the main text, for our analytic calculations we will consider $J_1$ as the overall energy scale, and set $J_1 = 1$.
We define the scaling relationships with respect to the multi-critical point at $(m_L, \mu_{\text{eff}}) = (0, 0)$, where $|k_0| \sim |\bs k|^4$.
Therefore, we choose  the following scaling relationships for a $d+1$ dimensional version of $S_{\text{eff}}$, with $d$ being the number of spatial dimensions, such that  the non-interacting part of $S_{\text{eff}}$ is invariant, 
\begin{align}
[k_0] = 4 [k_j] = 4; \qquad 
[\Phi] = - (d+8)/4,
\end{align}
where a quantity $X(\ell)$ evolves as $X(\ell) = e^{[X]\ell} X(0)$ with $\ell = \ln{(\Lambda_0/\Lambda)}$ being the logarithmic length scale and increases with decreasing energy. 
Therefore, with respect to the non-interacting fixed 
\begin{align}
[m_L] = 2; \qquad [A] = 0; \qquad [\gamma] = 0; \qquad  [\mu_{\text{eff}}] = 4; \qquad [g] = 4 - d.
\end{align}
We see that the coupling $g$ is strongly relevant in $d=2$, but becomes a marginally relevant perturbation to the Gaussian fixed point in $d=4$.
Therefore, we use $\epsilon = 4 - d$ as a small parameter for controllably studying the impact of interactions.
We will eventually take $\eps \to 2$ to understand the corresponding physics in $d=2$.
In order to perform the $\eps$ expansion, we generalize $\veps(\bs k)$ to a $d = 2n$ dimensional dispersion as
\begin{align}
\veps(\bs k) \to \veps_d(\bs k_1, \bs k_2) = 
m_L\qty(|\bs k_1|^2 + |\bs k_2|^2) + A\cos{\gamma} \qty(|\bs k_1|^4 + |\bs k_2|^4) + 2 A\sin{\gamma} |\bs k_1|^2  |\bs k_2|^2,
\end{align}
where $\bs k_j$ are $n$-dimensional vectors.

\begin{figure}[!t]
\centering
\includegraphics[width=0.5\columnwidth]{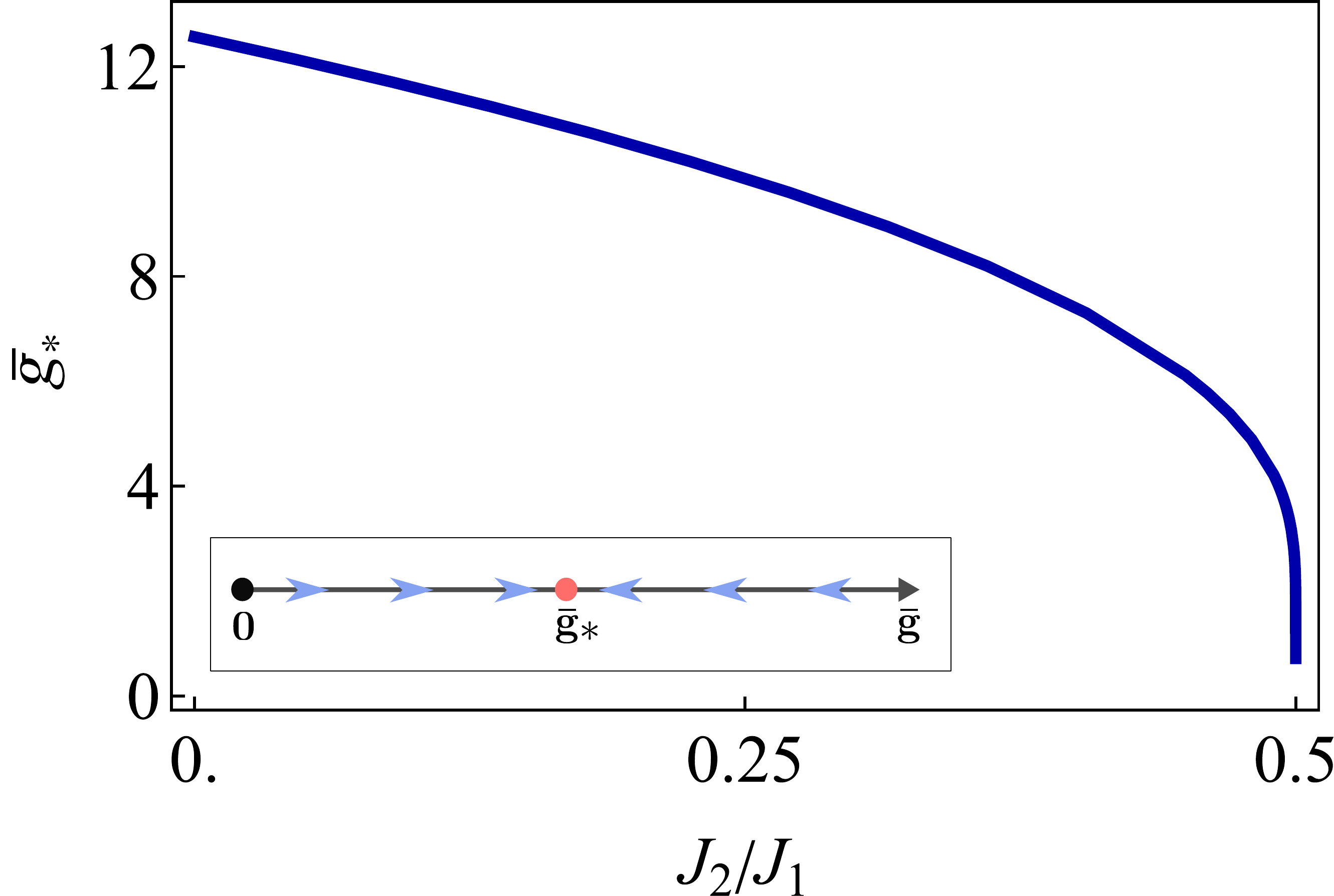}%
\hfill
\caption{Evolution of the interaction strength at the renormalization-group fixed point. 
Here, $\bar g_*$ is measured in units of $\eps = d - 4$ [see Eq.~(6) of the main text].
(Inset) Schematic representation of the renormalization group flow for the dimensionless coupling $\bar g$.
A non-trivial interacting fixed point is present for any $d < 4$.
}
\label{fig:g}
\end{figure}

Because of the short-ranged interaction among the bosons,  no self-energy correction is generated at one-loop.
The Hartree/Fock diagram renormalizes $\mu_{\text{eff}}$.
Since we are interested in the multi-critical point where $\mu_{\text{eff}} = 0$, a suitable bare $\mu_{\text{eff}}$ is chosen such that it cancels the quantum correction and the renormalized $\mu_{\text{eff}} = 0$. 
The amplitude for particle-hole excitations is controlled by
\begin{align}
\Gamma_{\text{ph}}(q) = g^2 \int \frac{\dd[d]{ k}}{(2\pi)^d} \frac{\dd{k_0}}{2\pi} G_0(k+q) G_0(k) 
=  g^2 \int \frac{\dd[d]{ k}}{(2\pi)^d} \frac{\dd{k_0}}{2\pi} \frac{1}{ik_0 + iq_0 + \veps_d(\bs k + \bs q) - \mu_{\text{eff}}} \frac{1}{ik_0 + \veps_d(\bs k) - \mu_{\text{eff}}}.
\end{align}
For $\Gamma_{\text{ph}}(q)$ to be finite the poles on the complex-$k_0$ plane must lie on opposite half planes.
This is no longer the case when $\mu_{\text{eff}} = 0$, and $\Gamma_{\text{ph}}(q)$ vanishes identically.
We note that this conclusion relies only on the positive semi-definiteness of $\veps_d(\bs k)$, and, thus, is independent of $d$.
Vertex correction in the particle-particle channel remains finite, however, and acquires the form
\begin{align}
\Gamma_{\text{pp}}(q = 0) =  g^2 \int \frac{\dd[d]{ k}}{(2\pi)^d} \frac{\dd{k_0}}{2\pi} G_0(k) G_0(-k) = \frac{g^2}{2} \int \frac{\dd[d]{ k}}{(2\pi)^d} \frac{1}{ 
m_L\qty(|\bs k_1|^2 + |\bs k_2|^2) + A\cos{\gamma} \qty(|\bs k_1|^4 + |\bs k_2|^4) + 2 A\sin{\gamma} |\bs k_1|^2  |\bs k_2|^2}.
\end{align}
We will evaluate the momentum intergral in $d=4-\epsilon$.
In order to do that we utilize the Hopf parameterization of the 3-sphere:
\begin{align}
\bs k_1 = k_r \mqty(
 \cos\zeta \cos\phi_1 \\
 \cos\zeta \sin\phi_1
), \qquad
\bs k_2 = k_r \mqty(
\sin\zeta \cos\phi_2 \\
 \sin\zeta \sin\phi_2
),
\end{align}
to obtain
\begin{align}
\Gamma_{\text{pp}}(q = 0) = \frac{g^2}{2(2\pi)^{4-\eps}} \int_0^{\pi/2} \dd{\zeta} \int_0^{2\pi}\dd{\phi_1} \dd{\phi_2} \int_{\Lam(1-\dl\ell)}^{\Lam} \dd{k_r} \frac{k_r^{3-\eps} \sin{\zeta} \cos{\zeta}}{ 
m_L k_r^2 + A\cos{\gamma} k_r^4 \qty(\cos^4{\zeta} + \sin^4{\zeta}) + 2A\sin{\gamma} k_r^4 \cos^2{\zeta} \sin^2{\zeta}}.
\end{align}
Because we are interested in the effects of interaction at the Lifshitz critical point, we set $m_L =0$ to obtain at leading order in $\eps$
\begin{align}
\Gamma_{\text{pp}}(q = 0) &= \frac{g^2}{4(2\pi)^{2} A} \underbrace{\int_0^{1} \dd{t} \frac{1}{\cos{\gamma} \qty[(1-t)^2 + t^2)] + 2\sin{\gamma}~ t(1-t) }}_{f_g(t)}. 
\end{align}
where we have introduced $t = \sin^2\zeta$.
The beta function for $g$ is presented in the main text, and the behavior of its  fixed point value is depicted in Fig.~\ref{fig:g}.

\section{Scaling behavior at finite temperatures and sub-critical fields} \label{sec:T}
In this section we demonstrate how the RG fixed point obtained in the previous section control scaling behaviors at finite temperatures ($T$) and sub-critical fields ($\mu_{\text{eff}} > 0$). 
Here, we will closely follow Ref.~\cite{sachdev1994}.
In this section, for notational convenience, we refer to $\mu_{\text{eff}}$ as $\mu$.

\subsection{$T = 0$ and $h < h_c$}
Here we provide a scaling estimate for the mean density of the bosons, which is equivalent to the magnetization. 
The density of bosons is related to the chemical potential and interaction through 
\begin{align}
\bar \rho_0(\ell) = \frac{\bar \mu(\ell)}{\bar g(\ell)}, \label{eq:rho}
\end{align}
where the notation $\bar X$ denotes the dimensionless part of the possibly dimensionful parameter $X$.
At $T=0$ and up to one loop order, the beta functions are ($z$ is the dynamical critical exponent)
\begin{align}
& \partial_\ell \bar\mu = z \bar \mu, \\
& \partial_\ell \bar g = (z-d) \bar g - C_d g^2, 
\end{align}
where the contribution from the Hartree/Fock diagram to $\partial_\ell \bar\mu$ is ignored because it does not add anything qualitatively new and can be compensated by a shift of the bare $\bar \mu$, and we have denoted the contribution from all vertex corrections as $C_d$.
As long as $|\bar \mu(\ell)| \ll 1$, $C_d$ is given by the particle-particle ladder diagram at leading order in $|\bar \mu(\ell)|$.
We allow the system to evolve  under the RG flow until $\bar \mu(\ell) \Lambda^z = \alpha \veps_\Lambda$, where  $\veps_\Lambda$ is the ultraviolet (UV) energy scale and $\alpha \ll 1$.
This allows us to define an RG `time', $\ell_*$, such that $\bar \mu(\ell_*) = \bar \mu(0) e^{z\ell_*} = \alpha \frac{\veps_\Lam}{\Lambda^z}$.
We assume $\alpha$ to be sufficiently large such that $\bar g(\ell_*) \approx \bar g_*$, its fixed point value.
Since $\rho_0$ has dimension of $\Lambda^d$,
\begin{align}
\bar \rho_0(0) = e^{- d \ell_*} \frac{\bar \mu(\ell_*)}{\bar g(\ell_*)} = e^{(z - d) \ell_*} \frac{\bar \mu(0)}{\bar g_*}.
\end{align}
Using the definition of $\ell_*$, we obtain
\begin{align}
\bar \rho_0(0) = \qty[\frac{\alpha \veps_\Lam}{\Lam^z }]^{1-d/z} \frac{ C_d \bar \mu(0)^{d/z}}{(z-d)  }.
\end{align}
Therefore, for the BEC criticality, where $z=2$, one obtains $\bar \rho_0(0) \sim \bar \mu(0)^{d/2}$~\cite{sachdev1994}, while for the Lifshitz criticality discussed here we obtain
\begin{align}
\bar \rho_0(0) \sim \bar \mu(0)^{d/4}.
\end{align}
This is reflected by the overall scaling in Eq. (8) of the main text.
When $m_L > 0$, a crossover behavior is present. 
In particular, in the presence of $m_L$ as $\mu$ increases, in the regime where $\mu \ll m_L \Lambda^2$ the scaling of $\bar \rho_0$ is controlled by the BEC fixed point and $\bar \rho_0 \sim \bar \mu^{d/2}$. 
At larger densities, $\mu \gg m_L \Lambda^2$, the system enters  the critical fan of the Lifshitz critical point, and the behavior of $\rho_0$ is controlled by the Lifshitz critical point, $\bar \rho_0 \sim \bar \mu^{d/4}$.

In section~\ref{sec:alg_liq} we will take a closer look at this regime, and show that a magnetic order in fact exists only when $m_L \neq 0$.

\subsection{$T > 0$}
We begin by noting that at a finite $T$ the quantum correction to $\mu$ from the Hartree/Fock diagram generates $T$ dependence, which is a qualitatively new feature~\cite{fisher1988, sachdev1994}.
Consequently, the beta function is modified as
\begin{align}
\partial_\ell \bar \mu = z \bar\mu - \frac{2 \Omega_d \bar g(\ell)}{\exp{\frac{\bar \varepsilon(\Lam) - \bar \mu }{ \bar T(\ell)}} - 1},
\end{align}
where $\Omega_d$ is the volume of the unit sphere in $d$-dimensions, and we have introduced $\bar \veps \equiv \veps/\Lam^z$.
Notice that $T$, being an energy scale, flows under RG,
\begin{align}
\partial_\ell \bar{T} = z \bar T \Rightarrow \bar T(\ell) = \bar T(0) e^{z\ell}.
\end{align}
Since we are working in a regime where $\mu \ll \veps(\Lambda)$, we ignore the $\bar \mu$ in the exponential factor and integrate over RG time to obtain
\begin{align}
\bar \mu(\ell) = - 2\Omega_d e^{z\ell} \int_0^\ell \dd{t} e^{-2t} \frac{ \bar{g}(t)}{\exp{\frac{\bar \varepsilon(\Lam)}{\bar T(t)}} - 1}.
\end{align}
Assuming the RG time $\ell_*$ to be sufficiently large such that $\bar g$ flows to its fixed point value $\bar g_*$, we obtain
\begin{align}
\bar \mu(\ell_*) =  \frac{\Omega_d \bar T(0) \bar g_*}{\bar \veps(\Lambda)} e^{z\ell_*} \left. \ln\qty(1 + \frac{1}{x}) \right|_{\exp{\frac{\bar \veps(\Lambda)}{\bar T(0) \exp{z\ell_*}}} - 1}^{\exp{\frac{\bar \veps(\Lambda)}{\bar T(0)}} - 1} 
\approx - \Omega_d  \bar g_* \frac{\bar T(0) \exp{z\ell_*}}{\bar \veps(\Lambda)}  \exp{-\frac{\bar \veps(\Lambda)}{\bar T(0) \exp{z\ell_*}}}
\end{align}
Notice that even if $\mu$ was zero initially, a finite $\mu$ is generated through thermal fluctuations. 
We allow this flow to continue until $|\mu(\ell)| \lesssim \veps(\Lam)$, i.e. $\bar \mu(\ell_*) = - \alpha' \bar \veps(\Lam)$ with $\alpha' \ll 1$~\cite{sachdev1994}.
We solve for $\ell_*$ to obtain
\begin{align}
\exp{-z\ell_*} = \frac{\bar T(0)}{\bar \veps(\Lambda)} W_0\qty(\frac{\Om_d \bar g_*}{\alpha' \bar \veps(\Lam)}), \label{eq:l*}
\end{align}
where $W_0$ is the principal branch of the Lambert $W$ function.

We compute the $T$ dependence of density at the RG time $\ell_*$~\cite{sachdev1994} by utilizing the result in \eqref{eq:l*},
\begin{align}
\bar \rho_0(\bar T(0)) = e^{-d \ell_*} \int \frac{\dd[d]{k}}{(2\pi)^d} \frac{1}{\exp{\frac{\bar \veps(\bs k) - \bar \mu(\ell_*)}{\bar T(\ell_*)}} - 1}. \label{eq:rho-T_append}
\end{align}
In order to establish $T$-dependent crossover behavior of $\rho_0$ it is convenient to introduce the parameter $\zeta_T = \bar T(0)^{1/z} \xi_L$, where $\xi_L = \sqrt{A/m_L}$ is the correlation length associate with the Lifshitz transition. 
In particular, when $\zeta_T \gg 1$ the scaling is controlled by the Lifshitz critical point with $z=4$ and $\rho \sim T^{d/4}$.
In the opposite regime, $\zeta_T \ll 1$, the scaling is controlled by the BEC fixed point with $z=2$, and $\rho \sim T^{d/2}$.
This result is portrayed in Fig. 3 of the main text by numerically evaluating \eqref{eq:rho-T_append} at $d=2$.

\section{Emergent algebraic liquid} \label{sec:alg_liq}
In this section we derive the hydrodynamic theory that describes long wavelength fluctuations of $\Phi$ at $m_L = 0$ but $\mu_{\text{eff}} > 0$.
It can be  viewed as an effective field theory of an interacting dense Bose gas undergoing a Lifshitz transition.
We begin with the position space version of the $2+1$ dimensional action in \eqref{eq:effS},
\begin{align}
S_M = \int \dd{\tau} \dd{\bs r} [
\Phi^*(\tau, \bs r) \qty{
\partial_\tau - m_L (\partial_x^2 + \partial_y^2) + A \cos{\gamma} ~(\partial_x^4 + \partial_y^4) + 2A \sin{\gamma} ~\partial_x^2 \partial_y^2 - \mu_{\text{eff}}
} \Phi(\tau, \bs r) + g |\Phi(\tau, \bs r)|^4 
].
\label{eq:effS_r}
\end{align}
Next, we obtain a hydrodynamic description of the dynamics of $\Phi$ in terms of its phase ($\vtheta$) and density ($\varrho$) fluctuations, which are introduced as follows,
\begin{align}
\Phi(\tau, \bs r) = \sqrt{\rho_0 + \varrho(\tau, \bs r)} ~e^{i \vtheta(\tau, \bs r)}.
\end{align} 
We substitute for $\Phi$ in \eqref{eq:effS_r} and obtain an effective action in terms of the hydrodynamic modes,
\begin{align}
S_M \simeq \int \dd{\tau} \dd{\bs r} \qty[
i \varrho \partial_\tau \vtheta + m_L \rho_0 |\bs \grad \vtheta|^2
+ A \rho_0 \cos{\gamma} \qty{(\partial_x^2 \vtheta)^2 + (\partial_y^2 \vtheta)^2}
+ A \rho_0 \sin{\gamma} ~(\partial_x^2 \vtheta)(\partial_y^2 \vtheta)^2
+ g \varrho^2 
+ \rho_0 \qty(g \rho_0 - \mu_{\text{eff}})
],   
\end{align}
where the spacetime dependence of the field variables has been made implicit for notational convenience, and we have dropped terms that contain total or higher derivatives, along with $\order{|\bs \grad \varrho|^2/\rho_0^2}$ terms.
The average density is determined by the condition for vanishing of the last term, 
\begin{align}
\rho_0 = \frac{\mu_{\text{eff}}}{g} = \frac{\tilde h_c - \tilde h}{g}.
\label{eq:rho0}
\end{align}
When $|m_L| > 0$ the low energy scaling is controlled by the Bose-Einstein condensation (BEC) critical point for which $g$ is dimensionless and \eqref{eq:rho0} produces the correct scaling of $\rho_0$ with $(h_c - h)$.
As we established in section~\ref{sec:rg}, with respect to the Lifshitz critical point ($m_L = 0$), the coupling $g$ is dimensionful in $d=2$; consequently, it will be convenient to use its  dimensionless version, $\bar g \equiv g/\rho_0$.
In this case, \eqref{eq:rho0} implies $\rho_0 = \sqrt{(\tilde h_c - \tilde h)/\bar{g}}$, which is consistent with the universal scaling form that was obtained in section~\ref{sec:T}.

The existence of an off-diagonal long ranged order (ODLRO) for $\Phi$ is encoded in the propagator of $\vtheta$.
Therefore, we integrate out $\varrho$ (this is done exactly) to obtain the  effective action for phase fluctuations,
\begin{align}
S_\vtheta =  \int \dd{\tau} \dd{\bs r} \qty[
\frac{1}{4g}(\partial_\tau \vtheta)^2
+ m_L \rho_0 |\bs \grad \vtheta|^2
+ A \rho_0 \cos{\gamma} \qty{(\partial_x^2 \vtheta)^2 + (\partial_y^2 \vtheta)^2}
+ A \rho_0 \sin{\gamma} ~(\partial_x^2 \vtheta)(\partial_y^2 \vtheta)^2
].
\end{align}
We presented the momentum space version of the effective action in Eq. (10) of the main text.

\begin{figure}[!t]
\centering
\includegraphics[width=0.5\textwidth]{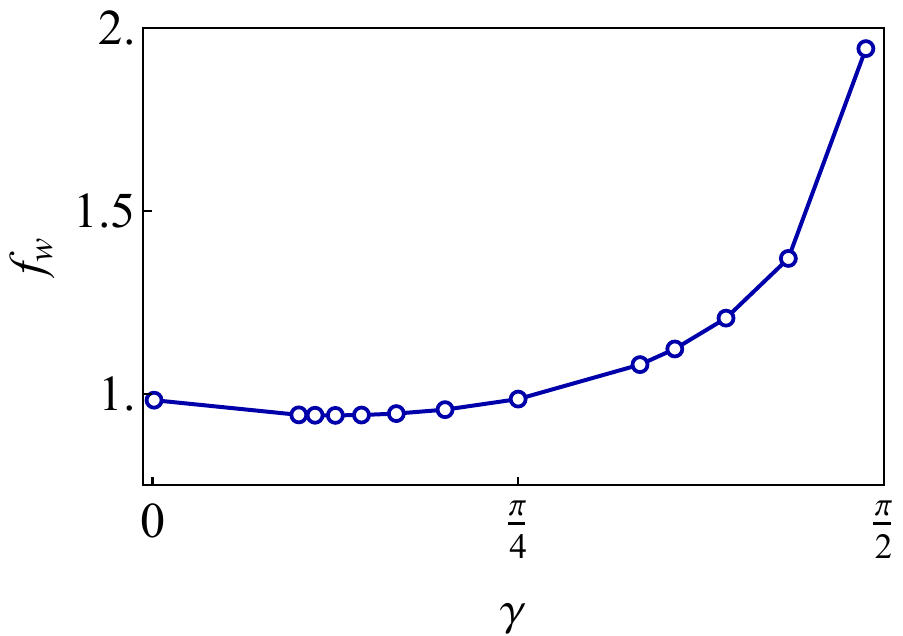}
\caption{The $\gamma$-dependence of the function  $f(\gamma)$ appearing in the exponent $\mc W(\gamma)$, scaled by a factor of $2^{3/4} \pi$.}
\label{fig:w}
\end{figure}

Existence of an ODLRO is determined by the equal-time correlation function,
\begin{align}
\avg{\Phi(0) \Phi^*(\bs r)} = \rho_0 e^{-\half \avg{\qty[\vtheta(0) - \vtheta(\bs r) ]^2}}.
\label{eq:corr-Phi}
\end{align}
In particular, if $\lim_{|\bs r| \to \infty} \avg{\Phi(0) \Phi^*(\bs r)} > 0$, then an ODLRO is present.
Therefore, existence of a Bose-Einstein condensation crucially relies on the finiteness of $\avg{\qty[\vtheta(0) - \vtheta(\bs r) ]^2}$ as $|\bs r| \to \infty$.
The key observation is that $\lim_{|\bs r| \to \infty} \avg{\qty[\vtheta(0) - \vtheta(\bs r) ]^2} < \infty$ only when $m_L > 0$.
The correlator is explicitly evaluated as follows,
\begin{align}
\avg{\qty[\vtheta(0) - \vtheta(\bs r) ]^2} &= \frac{2}{\rho_0} \int \dd{k} \frac{1 - \cos{(\bs k \cdot \bs r)}}{\frac{k_0^2}{4g \rho_0} + m_L |\bs k|^2 + A\cos\gamma (k_x^4 + k_y^4) + 2A \sin\gamma k_x^2 k_y^2} \nn \\
&= 2 \sqrt{\frac{g}{\rho_0 A}} \int \dd{\bs k} \frac{1 - \cos{(\bs k \cdot \bs r)}}{\sqrt{ |\bs k|^2/\xi_L^2 + \cos\gamma (k_x^4 + k_y^4) + 2  \sin\gamma k_x^2 k_y^2}} \label{eq:1}\\
& \equiv 2 \Gamma(\bs r, \xi_L),
\end{align}
where we have introduced the Lifshitz correlation length, $\xi_L = \sqrt{A/m_L}$.
Deep in the N\'{e}el antiferromagnetic phase, $\xi_L \ll 1$, and we retain only the first term in the denomaniator of the integrand to obtain
\begin{align}
\lim_{\xi_L \ll 1} \Gamma(\bs r, \xi_L) = \sqrt{\frac{g}{\rho_0 m_L}} \int \dd{\bs k} \frac{1 - \cos{(\bs k \cdot \bs r)}}{|\bs k|} 
= \frac{1}{2\pi |\bs r|} \sqrt{\frac{g}{\rho_0 m_L}} \int_0^{\Lam|\bs r|} \dd{t} \qty[1 - J_0(t)]
\end{align}
which saturates as $(\Lambda |\bs r|) \to \infty$, indicating the presence of an ODLRO.
As $m_L \to 0$, the correlation length diverges and we drop the first term in the denominator of the integrand in \eqref{eq:1} to obtain 
\begin{align}
\lim_{\xi_L \to \infty} \Gamma(\bs r, \xi_L) = \sqrt{\frac{g}{\rho_0 A}} \int \dd{\bs k} \frac{1 - \cos{(\bs k \cdot \bs r)}}{\sqrt{\cos\gamma (k_x^4 + k_y^4) + 2  \sin\gamma k_x^2 k_y^2}}.
\end{align}
It can be evaluated analytically at $\gamma = \pi/4$,
\begin{align}
\left. \lim_{\xi_L \to \infty} \Gamma(\bs r, \xi_L) \right|_{\gamma = \pi/4} = \sqrt{\frac{\sqrt{2} g}{\rho_0 A}} \int \dd{\bs k} \frac{1 - \cos{(\bs k \cdot \bs r)}}{|\bs k|^2}
= \frac{1}{2^{3/4}\pi} \sqrt{\frac{ g}{\rho_0 A}} \int_0^{\Lambda |\bs r|} \dd{t} \frac{1 - J_0(t)}{t},
\end{align}
which yields
\begin{align}
\left. \lim_{(\Lambda |\bs r|) \gg 1} \lim_{\xi_L \to \infty} \Gamma(\bs r, \xi_L) \right|_{\gamma = \pi/4} 
= \frac{1}{2^{3/4}\pi}\sqrt{\frac{ g}{\rho_0 A}} \ln(\Lambda |\bs r|) + \order{(\Lambda |\bs r|)^0}.
\end{align}
Therefore, in this limit (setting the ultraviolet cutoff $\Lambda = \sqrt{\rho_0}$ allows us to write $g = \rho_0 \bar g$), at the leading order in   $\sqrt{\rho_0} |\bs r| \gg 1$, we obtain
\begin{align}
\avg{\Phi(0) \Phi^*(\bs r)} \simeq \frac{\rho_0}{(\sqrt{\rho_0} |\bs r|)^{\sqrt{\bar{g}/A}/(2^{3/4}\pi) } },
\end{align}
which indicates an absence of ODLRO.
For $\gamma \neq \pi/4$, we express the asymptotic form of $\Gamma$ as
\begin{align}
\lim_{(\Lambda |\bs r|) \gg 1} \lim_{\xi_L \to \infty} \Gamma(\bs r, \xi_L)  
= \sqrt{\frac{g}{\rho_0 A}} ~f_{w}(\gamma)  ~\ln(\Lambda |\bs r|) \equiv \mc{W}(\gamma)~\ln(\Lambda |\bs r|),
\label{eq:w}
\end{align}
such that 
\begin{align}
\avg{\Phi(0) \Phi^*(\bs r)} \simeq \frac{\rho_0}{(\sqrt{\rho_0} |\bs r|)^{\mc{W}(\gamma)} },
\end{align}
The variation of $f_w$ with $\gamma$ is shown in Fig.~\ref{fig:w}.

\subsection{Stability}
We assess the stability of the algebraic liquid state against multi-particle condensates by computing the susceptibility of vertices of the form,
\begin{align}
\delta S_n = \tilde \lambda_n \int \dd{\tau}\dd{\bs r} \qty[\Phi^n(\tau, \bs r) + \mbox{h.c.} ] 
\simeq \lambda_n \int \dd{\tau}\dd{\bs r} \cos\qty[n \vartheta(\tau, \bs r)].
\end{align}
with $n\geq 2$.
In order to compute the scaling dimension of $\lambda_n$, we first compute the correlation function,
\begin{align}
\avg{ \cos\qty[n \vartheta(0, 0)] \cos\qty[n \vartheta(0, \bs r)]} = 2 \exp{-\frac{n^2}{2} \langle \qty[\vartheta(0, 0) - \vartheta(0, \bs r)]^2\rangle} = 2 \exp{- n^2 \Gamma(\Lambda |\bs r|, \xi_L \to \infty)}.
\end{align}
where in the final step we used the results obtained while evaluating Eq.~\eqref{eq:corr-Phi}.
As discussed earlier,  $\Gamma(\Lambda |\bs r|, \xi_L \to \infty)$ is logarithmically divergent in the limit $\Lambda |\bs r| \to \infty$, and the coefficient is a $\gamma$-dependent function (c.f. Eq.~\ref{eq:w} and Fig.~\ref{fig:w}).
Thus, we use an asymptotic form of $\Gam$ to extract the scaling exponent,
\begin{align}
\avg{ \cos\qty[n \vartheta(0, 0)] \cos\qty[n \vartheta(0, \bs r)]} \simeq 2 (\Lambda |\bs r|)^{- n^2 \mc W(\gamma)}.
\end{align}
Therefore, the operator $\cos\qty[n \vartheta(\tau, \bs r)]$ has a momentum-scaling dimension $ \frac{1}{2} n^2 \mc W(\gamma)$, which implies that the tree-level beta-function of $\lambda_n$ at the algebraic liquid fixed point is given by (this relationship is obtained by requiring $\delta S_n$ be scale invariant)
\begin{align}
\partial_\ell \lambda_n = 4 -  \frac{1}{2} n^2 \mc W(\gamma).
\end{align}
The beta-function implies that the operator with $n=2$ has the largest scaling dimension, and it remains an irrelevant perturbation as long as 
\begin{align}
\mc W(\gamma) > 2 \equiv \sqrt{\frac{g}{\rho_0 A}} f_w(\gamma) > 2.
\end{align}
We note that for a sufficiently dilute system, this constraint is satisfied. 
Since $\rho_0 \sim \sqrt{\mu_{\text{eff}}} \propto \sqrt{h_c - h}$, the algebraic liquid state is stable against multi-particle Bose condensates for sufficiently strong magnetic fields.

\section{Tensor Network Analysis}

In this section, we describe the numerical method used in this work, the infinite projected entangled-pair states (iPEPS), and also the detailed analysis of the numerical data. 

\subsection{iPEPS wavefunction}

In this work, we have used the iPEPS method~\cite{ipeps,pepstorch}, which was used to study the zero-field phase diagram of the $S=\frac{1}{2}$ $J_1-J_2-J_3$ model on the square lattice~\cite{liu2021}. In iPEPS, the wavefunction on the square lattice is expressed in terms of a product of tensors, on which the contraction over all the auxiliary indices is performed. Based on the insight from the linear spin wave theory (LSWT) results and preliminary density matrix renormalization group (DMRG) calculations, we choose a bipartite tiling of a $2\times 1$ unit cell containing two distinct tensors $a$ and $b$ which are to be optimized variationally based on energy~\cite{peps-optimization}, as shown below. The physical indices that correspond to the $S=1/2$ degrees of freedom are represented by black vertical lines. The auxiliary indices are represented by gray bonds, with bond dimension $D$ that controls the accuracy of the ansatz.
\begin{equation}
    |\psi(a,b)\rangle = \vcenter{\hbox{\includegraphics[scale=0.5]{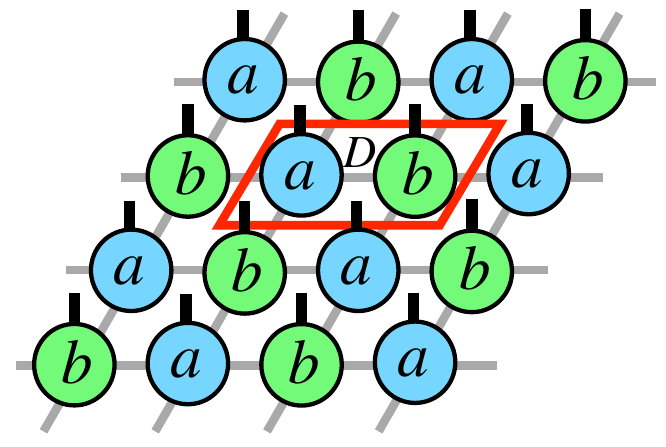}}}
    \label{eq:peps-ansatz}
\end{equation}

Such ansatz is capable of describing both the canted N\'eel antiferromagnetic states and potential liquid states whose in-plane components are isotropic. The evaluation of local observables such as energy and magnetization requires additional environment tensors, with environment bond dimension $\chi$ that controls the precision of the approximation for the environments. The environment tensors are constructed by the corner transfer matrix renormalization group (CTMRG) method~\cite{ctmrg98,corboz2014}. Typically, $\chi\geq D^2$ should be satisfied for a reasonable combination of control parameters.
\begin{figure}[tbh]
    \centering
    \includegraphics[scale=0.6]{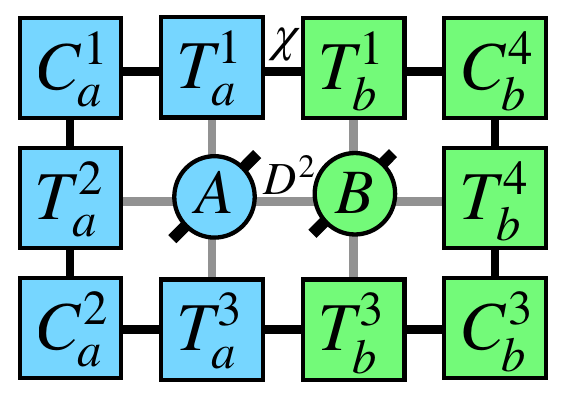},\quad
    \includegraphics[scale=0.6]{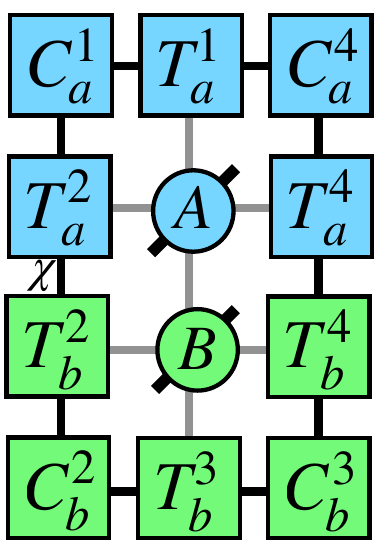}
    \caption{Double layers used for evaluation of 2-site operators.}
    \label{fig:envs}
\end{figure}

In our simulations, while we have pushed the bond dimensions up to $(D, \chi)=(7, 64)$ and $(D, \chi)=(6, 96)$, we find that local observables of concern have already achieved a very good convergence with $(D, \chi)=(5, 32)$. For example, the energy differences at $(\tilde{J_2},\tilde{J_3},h)=(1/4,1/8,3.97)$ are within $10^{-5}$ for $D$ ranging from $3$ to $7$ and $\chi$ from $32$ to $96$. Hence, we stick with this choice of control parameters to scan through the phase diagram.

\subsection{Fitting the uniform magnetization}

A variety of previous studies, analytical and numerical, suggest that the uniform magnetization of an $S=1/2$ Heisenberg antiferromagnet on square lattice shows a linear behavior with a logarithm correction up to the saturation field~\cite{sachdev1994,niku98-spinwave,lauc09}, in contrast to the root-singularity in one-dimensional systems~\cite{sachdev1994}. However, as demonstrated in the main context, our theory suggests that when approaching the critical lines, the boson density, which directly relates to the uniform magnetization in the spin language, also have a square root ($\mu=\frac{1}{2}$) scaling form with respect to the external field when close to the saturation field at $T=0$. Therefore, we expect the scaling behavior of the uniform magnetization up to the saturation field at $T=0$ to have a crossover from the linear form (with logarithm correction) to the square-root form.

To verify the existence of such a crossover, we fit our numerical data with the following function, which consists of two parts corresponding to the two distinct scaling forms for two-dimensional systems mentioned above. 
\begin{equation}
    [1/2 - \langle S^{(z)}\rangle] = \alpha_1 \Delta h \ln \frac{h_c}{\Delta h} + \alpha_2 \sqrt{\Delta h} \quad (\Delta h=h_c-h)
    \label{eq:fitting_func}
\end{equation}
where $h_c$ is the saturation field, and $\alpha_{1,2}$ give the relative weights of the two scaling functions.

\begin{figure}[thb]
    \centering
    \includegraphics[width=0.5\columnwidth]{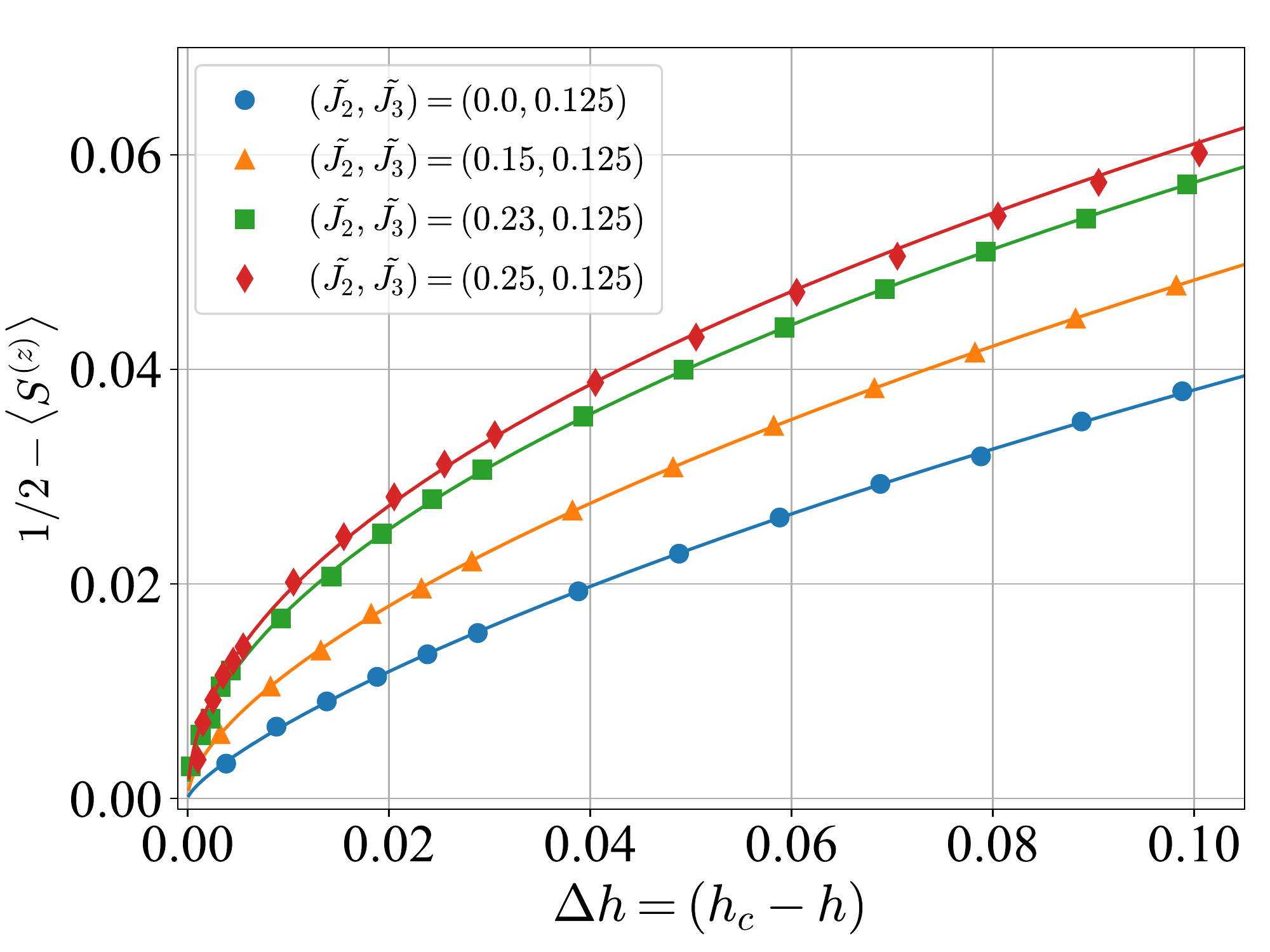}
    \caption{Fitting $D=5$ data with the universal fitting function, Eq.~\ref{eq:fitting_func}. 
    }
    \label{fig:fit_D=5}
\end{figure}

\twocolumngrid
\renewcommand{\emph}{\textit}
\bibliography{Non-BEC}


\end{document}